\documentclass[a4paper,11pt]{article}
\usepackage{jcappub} % for details on the use of the package, please see the JINST-author-manual
\usepackage[titletoc,toc,title]{appendix}

\usepackage{mathtools}
\usepackage{amsmath, amsfonts,bm}
\usepackage[utf8]{inputenc}
\usepackage{xcolor}
\usepackage{graphicx}
\usepackage{float}
\usepackage{subfig}

\usepackage{cleveref}

\newcommand\mathcomma{\,,}
\newcommand\mathperiod{\,.}

\DeclareMathAlphabet{\mathup}{OT1}{\familydefault}{m}{n}

\newcommand{\be}{\begin{equation}} 
\newcommand{\ee}{\end{equation}}

\usepackage{array}
\newcommand{\PreserveBackslash}[1]{\let\temp=\\#1\let\\=\temp}
\newcolumntype{C}[1]{>{\PreserveBackslash\centering}p{#1}}
\newcolumntype{R}[1]{>{\PreserveBackslash\raggedleft}p{#1}}
\newcolumntype{L}[1]{>{\PreserveBackslash\raggedright}p{#1}}

\title{Scalar field dark matter with time-varying equation of state}

\author[a]{Gaspard Poulot,}
\emailAdd{gmpoulot1@sheffield.ac.uk}
\affiliation[a]{School of Mathematics and Statistics, University of Sheffield, Hounsfield Road, Sheffield S3 7RH, United Kingdom}
\author[a,b]{Elsa M. Teixeira,}
\emailAdd{elsa.teixeira@umontpellier.fr}
\affiliation[b]{Laboratoire Univers \& Particules de Montpellier, CNRS \& Université de Montpellier (UMR-5299), 34095 Montpellier, France}
\author[a]{Carsten van de Bruck,}
\emailAdd{c.vandebruck@sheffield.ac.uk}
\author[c]{and Nelson J. Nunes}
\emailAdd{njnunes@ciencias.ulisboa.pt}
\affiliation[c]{Instituto de Astrof\'isica e Ci\^encias do Espa\c{c}o, Faculdade de Ci\^encias da Universidade de Lisboa, Campo Grande, Edif\'icio C8, P-1749-016, Lisboa, Portugal}

\date{}

\abstract{
We propose a new model of scalar field dark matter interacting with dark energy. Adopting a fluid description of the dark matter field in the regime of rapid oscillations, we find that the equation of state for dark matter is non-zero and even becomes increasingly negative at late times during dark energy domination. Furthermore, the speed of sound of dark matter is non-vanishing at all length scales, and a non-adiabatic pressure contribution arises. 
The results indicate that there are still unexplored possible interactions within the dark sector that lead to novel background effects and can impact structure formation processes. }

\begin{document}
\maketitle
\flushbottom

%\noindent\keywords{keyword 1; keyword 2; lower case except names, max 6 }\\

%\noindent\authorroles{For determining author roles, please use following taxonomy: \url{https://casrai.org/credit/}. Please list the roles for each author.} 

\section{Introduction}

The widely accepted standard cosmological model, known as $\Lambda$CDM, standing for ``a cosmological constant ($\Lambda$) plus cold dark matter (CDM)", is in impressive agreement with various observational findings. This framework explains the accelerated expansion of the Universe at late times, structure formation in the Universe and the properties of the cosmic microwave background radiation (CMB) (see, \textit{e.g.} \cite{Peebles:2022bya} for a recent overview). However, despite its success in explaining phenomena at large scales, the $\Lambda$CDM model still faces significant observational challenges when applied to clustering on smaller scales and consistently reconstructing the Universe's expansion history. In particular, the so-called Hubble tension is still a persistent challenge, together with other statistically less significant tensions. We refer to \cite{Abdalla:2022yfr} for an overview of these tensions and possible solutions. Moreover, from a theoretical perspective, the nature of the dark sector components, from which the model derives its name, remains a puzzle.

There is overwhelming observational evidence for dark matter (DM) from different cosmological and astrophysical probes,  including galaxy rotation curves \cite{1980ApJ...238..471R}, the peculiar motion of clusters \cite{1933AcHPh...6..110Z,1937ApJ....86..217Z}, the Bullet cluster system \cite{Clowe:2006eq}, observations from the CMB and baryonic acoustic oscillations \cite{Planck:2018vyg,2013ApJS..208...19H}. CDM constitutes approximately $26\%$ of the Universe's current energy density; therefore, illuminating its exact nature remains one of the most significant problems in cosmology and particle physics. A key factor is that dark matter must interact predominantly through gravity with the particles of the Standard Model of particle physics to assist structure formation in the Universe. The lack of direct detection of weakly-interacting massive particles (WIMPs) \cite{Gottel:2024cfj,XENON:2018voc,LUX:2016ggv,Kahlhoefer:2017dnp,Bertone:2018krk,Graham:2015ouw,Choi:2020rgn,Amruth:2023xqj}, the simplest and most popular CDM candidates, motivates the search for alternative proposals rooted in theories of high-energy particle physics. 
One such proposal of relevance for this work is the concept of \textit{fuzzy dark matter} \cite{Hu:2000ke,Dentler:2021zij,Svrcek:2006yi,Arvanitaki:2009fg,Hlozek:2014lca}, which consists of modelling dark matter as an ultralight scalar field or axion-like particle with a mass around $10^{-22}$ eV. The ultra-light nature of these particles translates into a large \textit{de Broglie} wavelength, effectively suppressing structure formation on small scales, while retaining the successes of the CDM paradigm on larger scales \cite{Bullock:2017xww,Weinberg:2013aya,Hui:2016ltb,Rogers:2023ezo,Lague:2020htq}. Ultralight scalars often emerge in particle physics and string theory compactifications, where axion-like particles arise as Kaluza-Klein zero modes of anti-symmetric tensor fields \cite{Green:1987sp,Svrcek:2006yi,Arvanitaki:2009fg}. Arguments based on measurements of CMB anisotropies constrain the mass of these ultralight fields to be $m_{\phi} \gtrsim 10^{-24}$ eV \cite{Amendola:2005ad,Hlozek:2016lzm,Hlozek:2017zzf,Hlozek:2014lca,Farren:2021jcd}, while observations of the Lyman$-\alpha$ forest extend this lower bound to about $m_{\phi} \sim 10^{-21}$ eV \cite{Irsic:2017yje,Armengaud:2017nkf,Kobayashi:2017jcf,Nori:2018pka,Rogers:2020ltq}, assuming these particles account for over $30\%$ of the total dark matter content. Albeit less widely agreed upon, studies on the kinematics of ultra-faint dwarf galaxies further bring the lower limit to around $m_{\phi} \sim 10^{-19}$ eV \cite{Hayashi:2021xxu, Dalal:2022rmp,Goldstein:2022pxu}. Other probes include galaxy clustering \cite{Lague:2021frh,Rogers:2023ezo}, weak lensing measurements \cite{Dentler:2021zij,Kunkel:2022ldl} and 21 cm observations \cite{Hotinli:2021vxg,Bauer:2020zsj,Flitter:2022pzf}. Furthermore, recent pulsar timing array reports of a stochastic gravitational wave background impose significant constraints on ultralight axions \cite{EPTA:2023xxk, NANOGrav:2023gor}. It has been shown that the travel time of pulsar radio beams is influenced by the gravitational potential induced by such ultralight DM particles \cite{Khmelnitsky:2013lxt,Porayko:2014rfa,Porayko:2018sfa,Xia:2023hov}.  
The analysis of the second data release from the European Pulsar Timing Array \cite{EuropeanPulsarTimingArray:2023egv} peaks at $m_{\phi} \sim 10^{-23}$ eV and rules out particles with masses ranging from $10^{-24}$ eV to $10^{-23.3}$ eV. It should be noted that the analyses mentioned above often rely on the assumption that the ultralight DM makes up the whole of DM and interacts solely through gravitational means.

The other dark ingredient of the standard model of cosmology is the cosmological constant $\Lambda$, the simplest realisation of dark energy in the form of a background energy component which accounts for the current accelerated expansion of the Universe. Given that the value of the cosmological constant has to be very small, alternative candidates for DE have been proposed, including slowly evolving scalar fields \cite{Wetterich:1987fm,Peebles:1987ek,Ratra:1987rm,Wetterich:1994bg,Caldwell:1997ii}, three-form fields \cite{Koivisto_2013} and other more exotic proposals (see \cite{Amendola:2015ksp,Copeland:2006wr,Li:2011sd} for reviews on dark energy models). If DE is a dynamical degree of freedom, couplings to other matter forms are generally expected unless a particular symmetry forbids such interactions \cite{Carroll:1998zi}. Constraints coming from the \textit{Cassini} probe \cite{Bertotti:2003rm} suggest that the coupling of a slowly rolling DE scalar field to the standard model fields has to be much weaker than gravity, as the field's minute mass results in a very large interaction range. On the other hand, these constraints can be relaxed for couplings to DM only, derived based on model-dependent cosmological observations. Some interacting dark energy (IDE) models may address shortcomings of the $\Lambda$CDM model such as the Hubble tension \cite{Abdalla:2022yfr,Schoneberg:2021qvd,DiValentino:2021izs,DiValentino:2017iww,DiValentino:2019ffd,Yang:2021hxg}, providing a guiding direction and framework for further study. Of particular interest are a class of models in which the CDM-DE coupling is directly proportional to the energy density of dark energy, as studied for example in Refs.~\cite{Honorez_2010,Gavela:2010tm,He:2008si,Valiviita:2008iv,Yang:2017ccc,Gavela:2009cy,Yang:2022csz,Yang:2019uzo,Yang:2020uga,Nunes:2022bhn,vandeBruck:2022xbk,DiValentino:2019jae,Giare:2024ytc}.

In this paper, we propose a model of scalar field DM, which interacts with a quintessence-like DE field. A conformal coupling between the DM and the DE field describes this interaction. In the literature, this interaction is mediated by the slowly evolving scalar degree of freedom identified with DE \cite{Carrillo_Gonz_lez_2018,Johnson:2020gzn}. In this work, however, we propose an alternative formulation with the twist that the coupling depends instead on the (fast-oscillating) DM scalar degree of freedom. We note here that even if the DE sector is considered here to be a slowly evolving scalar field, we expect many of our results to still hold if DE is a cosmological constant (but conformally coupled to DM) or a three-form field \cite{Koivisto_2013}. Nevertheless, our formalism is not circumscribed to the conformal coupling. In Appendix B we present an alternative formulation of a theory with a DM-DE coupling in which DM has a very similar phenomenology. 

This article is organised as follows. In \cref{sec:model}, we motivate and describe the proposed framework. In this section, we also collect the field equations and useful relations, which will be relevant to the subsequent calculations. In \cref{sec:fluid}, we derive an effective fluid-field description, a valid approximation from the onset of the DM field quick oscillations around the minimum of an effective potential. The treatment of the associated fluid-field cosmological perturbations is discussed in \cref{sec:pert}. We summarise our findings in \cref{sec:conc}. 

\section{Model} \label{sec:model}

Our setup is inspired by that of field theories of dark energy, such as coupled quintessence \cite{Wetterich:1994bg,Amendola:1999er,Carrillo_Gonz_lez_2018}, but with the role of the fields for DM and DE swapped. That is, we treat $\phi$ in the action below as a dark matter scalar field, and the dark energy sector is coupled conformally to the DM sector, resulting in the following effective action for this model:
\begin{eqnarray}\label{eq:new_coupledquintessence_action}
{\cal S} = \int d^4 x \sqrt{-g}\left( \frac{M_{\rm Pl}^2}{2} {\cal R} - \frac{1}{2}g^{\mu\nu} \partial_\mu \phi \partial_\nu\phi - U(\phi) \right) + {\cal S}_{\rm SM} + {\cal S}_{\rm DE} \mathcomma
\end{eqnarray}
where ${\cal R}$ is the Ricci-scalar and $M_{\rm Pl} \equiv 1/\sqrt{8 \pi G}$ is the reduced Planck mass. In the action above, ${\cal S}_{\rm SM}$ denotes the Lagrangian containing the standard model fields propagating in geodesics of the metric $g_{\mu\nu}$. In what follows, we assume that the DE sector can be described by a slowly evolving scalar field $\chi$, described by the action  
\begin{equation}\label{eq:DEaction}
{\cal S}_{\rm DE} = \int d^4 x \sqrt{-{\tilde g}} \left( \frac{1}{2}\tilde g^{\mu\nu}\partial_\mu \chi \partial_\nu \chi - V(\chi) \right)~,
\end{equation}
where the metric ${\tilde g}$ is related to the metric $g$ via a conformal transformation of the form ${\tilde g}_{\mu\nu} = C(\phi)g_{\mu\nu}$. Rewriting ${\cal S}_{\rm DE}$ in terms of the metric $g_{\mu\nu}$ results in 
\begin{equation}\label{eq:DEaction2}
{\cal S}_{\rm DE} = \int d^4 x \sqrt{-g} \left( \frac{C(\phi)}{2}g^{\mu\nu}\partial_\mu \chi \partial_\nu \chi - C^2(\phi) V(\chi) \right) \mathperiod
\end{equation}
In contrast to models such as coupled quintessence, in which $C$ depends on the dark energy field, the function $C$ in this framework is dependent on {\it dark matter} properties\footnote{In Appendix A we show the corresponding equations for the case of a conformal coupling depending also on the kinetic energy of the DM field. However, this is a different model and its discussion is out of the scope of this work.}. 

The equation of motion for the DE field $\chi$ is derived from the corresponding variation of the action and reads
\begin{equation}\label{eq:KGequation}
\nabla^\mu \nabla_\mu  \chi - C \frac{dV}{d\chi} = - \frac{\nabla_\alpha C}{C}\nabla^\alpha \chi  \mathperiod
\end{equation}
The equation for the DM field $\phi$ is derived in an analogous manner and yields  
\begin{equation} \label{eq:KGequation2}
\nabla^\mu \nabla_\mu \phi - \frac{\partial U}{\partial \phi} = Q \mathcomma
\end{equation}
where the coupling $Q$ was defined as
\begin{eqnarray}\label{eq:coupling}
Q &=& \nabla_\mu\left( \frac{\partial L_\chi}{\partial(\nabla_\mu \phi)} \right) - \frac{\partial L_\chi}{\partial \phi} \nonumber \\
&=&  \frac{C_{,\phi}}{2C}\left[ Cg^{\alpha\beta}\nabla_\alpha\chi \nabla_\beta \chi + 4C^2U  \right] \mathperiod
\end{eqnarray}

The Einstein equations result from variation of the action with the gravitational metric $g_{\mu \nu}$, resulting in
\begin{equation}\label{eq:eineq}
    R_{\mu \nu} - \frac{1}{2} R g_{\mu \nu} = \kappa^2 \left( T_{\mu\nu}^{(\phi)} + T_{\mu\nu}^{(\chi)} + T_{\mu\nu}^{(\rm SM)} \right) \mathcomma
\end{equation}
where $R_{\mu \nu}$ is the Ricci tensor, $\kappa \equiv 1/ M_{\rm Pl}$ and $T_{\mu\nu}^{(i)}$ are the energy-momentum tensors for each $i$-th fluid, which for the dark sector scalar fields $\phi$ and $\chi$ read 
\begin{eqnarray}
T_{\mu\nu}^{(\phi)} = \nabla_\mu \phi \nabla_\nu \phi - g_{\mu\nu} \left( \frac{1}{2}g^{\alpha\beta} \nabla_\alpha\phi\nabla_\beta \phi + V(\phi) \right) \mathcomma
\end{eqnarray}
\begin{eqnarray}
T_{\mu\nu}^{(\chi)} &=& C \nabla_\mu \chi \nabla_\nu \chi - g_{\mu\nu} \left( \frac{C}{2}g^{\alpha\beta} \nabla_\alpha\chi\nabla_\beta \chi + C^2 U(\chi) \right) \mathcomma
\end{eqnarray}
respectively, and obey the following conservation equations:
\begin{equation}
   \nabla^\mu T_{\mu\nu}^{(\phi)} = Q \nabla_\nu \phi \mathcomma \quad \nabla^\mu T_{\mu\nu}^{(\chi)} = -Q \nabla_\nu \phi \mathcomma
\end{equation}
which are physically equivalent to the Klein-Gordon equations. 

For the background cosmology we consider the case of a flat Friedmann-Lemaître-Robertson-Walker Universe, for which the field equations read
\begin{equation}
    \ddot{\phi} + 3H \dot{\phi} + U_{,\phi} = \frac{1}{2} C_{,\phi} \dot{\chi}^2 - 2 C C_{, \phi} V \mathcomma
\end{equation}
\begin{equation}\label{eq:chi_KG}
    \ddot{\chi} + \left( 3 H + \frac{C_{,\phi}}{C} \dot{\phi} \right) \dot{\chi} + C V_{,\chi} = 0 \mathcomma
\end{equation}
where $H$ is the Hubble expansion rate and over-dots indicate derivatives according to cosmic time $t$.
The energy densities $\rho_\phi$ and $\rho_\chi$ and the corresponding pressures $p_\phi$ and $p_\chi$ are given by 
\begin{equation}
    \rho_{\phi} = \frac{1}{2} \dot{\phi}^2 + U(\phi),\ \ \ p_{\phi} = \frac{1}{2} \dot{\phi}^2 - U(\phi) \mathcomma
\end{equation}
\begin{equation}
    \rho_{\chi} = \frac{1}{2} C \dot{\chi}^2 + C^2 V(\chi),\ \ \ p_{\chi} = \frac{1}{2} C \dot{\chi}^2 - C^2 V(\chi) \mathperiod
\end{equation}
The Friedmann equations, which govern the expansion of the Universe and which can be derived from Einstein's equations, Eq.~\eqref{eq:eineq}, read 
\begin{equation}
    H^2 = \frac{\kappa^2}{3} \left[ \frac{1}{2} \dot{\phi}^2 + U(\phi) + \frac{1}{2} C(\phi) \dot{\chi}^2 + C(\phi)^2 V(\chi) \right] \mathcomma
\end{equation}
\begin{equation}
    \dot{H} = - \frac{\kappa^2}{3} \left[ \dot{\phi}^2 + C(\phi) \dot{\chi}^2 \right] \mathperiod
\end{equation}
Finally, the Klein-Gordon equations for $\phi$ and $\chi$, Eqs.~\eqref{eq:KGequation} and \eqref{eq:KGequation}, become
\begin{equation}\label{eq:phi_cons}
    \dot{\rho}_{\phi} + 3H \left( \rho_{\phi} + p_{\phi} \right) = - \frac{C_{,\phi}}{2C} \left( \rho_{\chi} - 3 p_{\chi} \right) \dot{\phi} \mathcomma
\end{equation}
\begin{equation}
    \dot{\rho}_{\chi} + 3H \left( \rho_{\chi} + p_{\chi} \right) =  \frac{C_{,\phi}}{2C} \left( \rho_{\chi} - 3 p_{\chi} \right) \dot{\phi} \mathperiod
\end{equation}

Since we wish for $\phi$ to behave as dark matter and $\chi$ as dark energy, in this analysis we focus on the following self-interacting scalar field potentials:
\begin{equation}\label{eq:vp}
    V (\chi) = V_0 e^{- \kappa \lambda \chi} \mathcomma
\end{equation}
\begin{equation} \label{eq:up}
    U (\phi) = \frac{1}{2} m^2 \phi^2 \mathperiod
\end{equation}
The function $Q$ defined in Eq.~\eqref{eq:coupling} can be written as
\begin{equation} \label{eq:couplingp}
    Q= \frac{C_{,\phi}}{2C} \left( \rho_{\chi} - 3 p_{\chi} \right) \mathcomma
\end{equation}
for a field dependent conformal function $C$.
Unlike standard interacting dark energy-dark matter models, we emphasise that the coupling in the present model depends on the DE energy density and pressure, rather than the DM density. 

This concludes the derivation of the relevant equations for this theory, which will be used in the next section to derive a fluid-field description for the system. 

\section{Towards a fluid-field description} \label{sec:fluid}

We now seek to solve the equations derived in the previous section to study the evolution of both the DM and DE species. The main difference and challenge of our model compared to interacting quintessence models is that the coupling described here by $Q(t)$ depends on $\phi$ and $\dot\phi$, which must be rapidly oscillating for $\phi$ to behave like DM.  Therefore, we have to perform an analysis based on the time-averaged character of the field over a period of oscillation to extract its average evolution and make the equations more manageable. For the particular conditions under study, the Klein-Gordon equation for $\phi$ reads

\begin{equation}\label{eq:kgeqQ}
    \ddot{\phi} + 3H \dot{\phi} +  m^2\phi = -Q(t) \mathperiod
\end{equation}
We now consider a conformal coupling function linearly dependent on $\phi$, \textit{i.e.} $C = 1+2\kappa\beta \phi$, where $\beta$ is a constant and $\kappa = 1/M_{\rm Pl}$. This allows us to decompose the source term in Eq.~\eqref{eq:kgeqQ} as
\begin{equation}
    Q(t)=Q_0(t)+Q_1(t) \phi \mathcomma
\end{equation}
with 
\begin{eqnarray}
Q_0(t) \equiv -\kappa\beta (\dot{\chi}^2-4V) \mathcomma \hspace{2cm}
Q_1(t) \equiv  8\kappa^2\beta^2 V \mathcomma
\end{eqnarray}
%$Q_0 \equiv -\kappa\beta (\dot{\chi}^2-4V)$ and $Q_1 \equiv -2 \kappa^2\beta^2 (\dot{\chi}^2-8V).$ 
%
such that the equation of motion for $\phi$ can be recast into
\begin{equation}\label{eomapproxphi}
     \ddot{\phi} + 3H \dot{\phi} +  m^2_{\rm eff}(t)\phi = -Q_0(t) \mathcomma
\end{equation}
where we have defined the effective mass 
\begin{eqnarray} \label{eq:meff}
m^2_{\rm eff}(t)\equiv m^2+Q_1(t) \mathperiod    
\end{eqnarray}
Before solving these equations, we want to stress that the following analysis hinges on the form of Eq.~\eqref{eomapproxphi} and any theory which predicts this form for the equation of motion for the DM field, will result in a similar DM phenomenology, as long as the functions $Q_0$ and $Q_1$ are slowly varying (and it's not even a necessary condition that these quantities must depend on the dark energy scalar field). What will potentially change is the dynamics of the DE field. We give another example of such a theory in Appendix B. 

We will be interested in the case in which $m>H$ and the scalar field is under-damped, allowing for the oscillations to begin before the DM dominated epoch begins. To solve Eq.~\eqref{eomapproxphi} we employ an ansatz for $\phi$:
\begin{eqnarray}
    \phi(t)=\phi_{\rm osc}(t)+A(t) \mathcomma
\end{eqnarray}
where $\phi_{\rm osc}$ is the solution to the homogeneous version of the Klein-Gordon equation: 
\begin{equation}
    \ddot{\phi}_{\rm osc} + 3H \dot{\phi}_{\rm osc} +  m^2_{\rm eff}(t)\phi_{\rm osc} = 0 \mathcomma
\end{equation}
while $A(t)$ solves 
\begin{equation}
    \ddot{A} + 3H \dot{A} + m^2_{\rm eff}(t)A =-Q_0(t) \mathperiod
\end{equation}
On the basis that $A$ is sourced by a slowly-evolving function, $Q_0$, we make the assumption that $A(t)$ itself is slowly varying such that 
$\ddot{A}, 3H\dot{A} \ll m^2_{\rm eff}A$ and thus, we can neglect the first two terms of the above equation. This results in an approximated analytical solution for $A$: 
\begin{eqnarray} \label{eq:at}
    A(t) \approx -\frac{Q_0(t)}{m^2_{\rm eff}(t)} \mathperiod
\end{eqnarray}
Making use of the Wentzel-Kramer-Brillouin (WKB) approximation, we find the following solution for $\phi_{\rm osc}$:
\begin{equation}
    \phi_{\rm osc}(t)=\left( \frac{a_0}{a} \right)^{3/2}\left(\frac{m_0}{m_{\rm eff}} \right)^{1/2}\left( \phi_+ \sin{(m_{\rm eff} t) }+\phi_- \cos{(m_{\rm eff} t)}\right) \mathcomma
\end{equation}
where $m_0$ is the effective mass taken at $t=t_0$, and $\phi_+$ and $\phi_-$ are constants. We can then average $\phi$ over one oscillation period to get
\begin{equation}\label{phi}
\langle \phi \rangle=A \mathcomma
\end{equation}
and for its velocity
\begin{equation}\label{phidot}
\langle \dot{\phi} \rangle=\dot{A} \mathperiod
\end{equation}
Likewise, the variance of these quantities becomes,
\begin{equation}\label{eq:phi2variance}
    \langle \phi^2\rangle=\frac{1}{2}(\phi_+^2+\phi_-^2) \left(\frac{m_0}{m_{\rm eff}}\right)\left( \frac{a_0}{a} \right)^3 + A^2 \mathcomma
\end{equation}
\begin{equation}
      \langle \dot{\phi}^2\rangle=\frac{1}{2}m_{\rm eff}^2 (\phi_+^2+\phi_-^2) \left(\frac{m_0}{m_{\rm eff}}\right)\left( \frac{a_0}{a} \right)^3 \mathcomma
\end{equation}
where in the last line we have ignored an $\dot{A}^2$ term, since under our assumptions
$\dot A \ll m_{\rm eff} A$.
%we assume $m_{\rm eff} A \gg \dot A$.
%

\begin{figure}[h]
    \centering
    \includegraphics[height=0.6\linewidth]{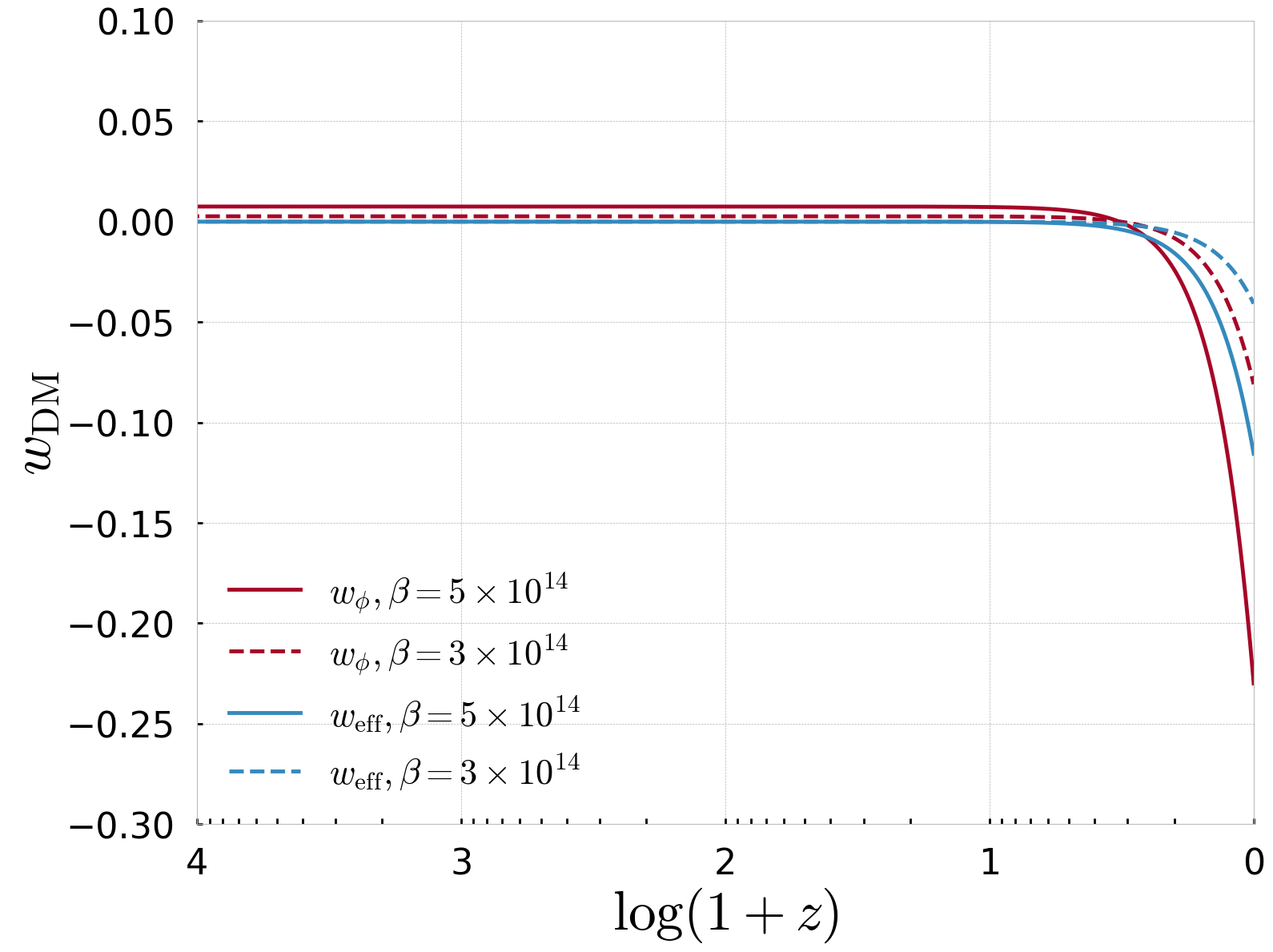}
    \caption{Evolution of the equation of state of dark matter $w_{\phi}$ (red curves) and the effective dark matter equation of state $w_{\rm eff}$ (blue curves), defined in Eq.~\eqref{wphi} and Eq.~\eqref{eq:weff} respectively. The model parameters used in this illustrative example are $m=10^{-17}$ eV and $\lambda=-0.2$, according to Eqs.~\eqref{eq:vp} and \eqref{eq:up}. Filled lines label the curves with $\beta = 5 \times 10^{14}$, while dashed lines are used for $\beta = 3 \times 10^{14}$.}
    \label{fig:wDM}
\end{figure}

The above expressions are crucial to the final results shown in this section. For standard uncoupled scalar field DM, both the field value and its derivative average out to zero. These equations encode the fact that, in our case, the field does not oscillate around zero, but rather, its averaged value is shifted by $A$ due to the interaction. Therefore, the average energy density and pressure of the field are
\begin{equation}\label{avenergy}
    \langle \rho_{\phi} \rangle =\frac{1}{4}(\phi_+^2+\phi_-^2) \left(\frac{m_0}{m_{\rm eff}}\right)\left( \frac{a_0}{a} \right)^3 (m_{\rm eff}^2+m^2)+ \frac{m^2A^2}{2} \mathcomma
\end{equation}
and
\begin{equation}
    \langle p_{\phi}\rangle=\frac{1}{4}(\phi_+^2+\phi_-^2) \left(\frac{m_0}{m_{\rm eff}}\right)\left( \frac{a_0}{a} \right)^3 (m_{\rm eff}^2-m^2)- \frac{m^2A^2}{2} \mathcomma
\end{equation}
respectively. Here we see clearly that the interaction between DM and DE results in a change in the DM pressure, causing its departure from zero. 
The pressure can also be expressed as
\begin{equation}
     \langle p_{\phi}\rangle= \left( \langle \rho_{\phi} \rangle -\frac{m^2A^2}{2} \right)\frac{m^2_{\rm eff}-m^2}{m^2_{\rm eff}+m^2} -\frac{m^2A^2}{2} \mathcomma
\end{equation}
in which case the equation of state becomes
\begin{equation}
\label{wphi}
    w_{\phi}\equiv \frac{\langle p_{\phi}\rangle}{\langle \rho_{\phi} \rangle} = \left( 1 -\frac{m^2A^2}{2 \langle \rho_{\phi} \rangle} \right)\frac{m^2_{\rm eff}-m^2}{m^2_{\rm eff}+m^2} -\frac{m^2A^2}{2 \langle \rho_{\phi} \rangle} \mathperiod
\end{equation}
Provided the kinetic energy of the dark energy field is small comparatively to its potential, $Q_1$ is positive and $m_{\rm eff} > m$. This means that the equation of state of dark matter starts by being positive in its early stages. Around matter-dark energy equality, the equation of state begins to decrease and, in fact, becomes increasingly negative, as illustrated in Fig.~\ref{fig:wDM}. The very late-time occurrence of this transition is ascribed to the dependence of the coupling on the dark energy density, implying that the coupling only has a significant impact on $w_{\phi}$ when DE begins to dominate. Indeed, from Eq.~\eqref{wphi}, we see that the equation of state parameter for dark matter becomes negative when
\begin{equation} \label{eq:rhophib}
    \langle \rho_\phi \rangle < \frac{1+Z}{Z} \, \frac{m^2 A^2}{2}
\end{equation}
where $Z=(m^2_{\rm eff}-m^2)/(m^2_{\rm eff}+m^2) \approx Q_1/2m^2$. As $A\approx -Q_0/m^2$, it turns out that $w_\phi < 0$ when $\langle \rho_\phi \rangle$ drops below
\begin{equation}
    \langle \rho_\phi \rangle < \frac{Q_0^2}{Q_1} \approx V \approx \rho_\chi \,.
\end{equation}
Consequently, the DM pressure becomes appreciably negative only at very low redshift.

%%%%%%%%%%%%%%%%%%%%%%%
Differentiating the averaged density, Eq.~\eqref{avenergy}, we obtain 
\begin{equation}
    \frac{d\langle{\rho_{\phi}}\rangle}{dt}+3H \left( \langle \rho_{\phi} \rangle+\langle p_{\phi}\rangle \right)=\frac{3H}{4}(\phi_+^2+\phi_-^2) \left(\frac{m_0}{m_{\rm eff}}\right)\left( \frac{a_0}{a} \right)^3 (m_{\rm eff}^2-m^2)+m^2A\dot{A} \mathcomma
\end{equation}
or equivalently
\begin{equation}\label{eq:averagedfluideq}
    \frac{d}{dt}\langle{\rho_{\phi}}\rangle+3H \left( \langle \rho_{\phi} \rangle-\frac{m^2A^2}{2} \right)=m^2A\dot{A} \mathperiod
\end{equation}
Had we done the time-averaging of Eq.~\eqref{eq:phi_cons} directly, we would conclude that, in fact
\begin{eqnarray}
 \frac{d\langle{\rho_{\phi}}\rangle}{dt} = \langle{\dot\rho_{\phi}}\rangle   \mathperiod
\end{eqnarray}
We now define an effective equation of state parameter for DM as
\begin{equation} \label{eq:weff}
    w_{\rm eff} = -\frac{1}{\langle{\rho_{\phi}}\rangle}\left( \frac{m^2A^2}{2}+ \frac{m^2A\dot A}{3H} \right) \mathcomma
\end{equation}
such that Eq. \eqref{eq:averagedfluideq} becomes
\begin{equation}
    \langle{\dot\rho_{\phi}}\rangle + 3H \langle{\rho_{\phi}}\rangle (1+w_{\rm eff})=0 \mathperiod
\end{equation}
The qualitative behaviour of $w_{\rm eff}$ is shown in Fig.~\ref{fig:wDM}, derived from a modified version of the publicly available \texttt{CLASS}\footnote{\href{https://github.com/lesgourg/class_public}{https://github.com/lesgourg/class\_public}} \cite{Blas:2011rf} code to account for the dynamics of the present model. From this, we see that the DM component effectively evolves like DM during matter domination and then transitions into a negative pressure fluid behaviour during DE domination. 

Finally, and for completeness, the equation of motion for the dark energy field, obtained by substituting Eqs.~\eqref{phi} and \eqref{phidot} into Eq.~\eqref{eq:chi_KG}, yields
\begin{equation}
\ddot{\chi}+\left(3H + \frac{2\kappa\beta\dot A}{1+2\beta\kappa A} \right)\dot{\chi}+V_{,\chi}\left(1+2\beta\kappa A\right) = 0.
\end{equation}

Before discussing cosmological perturbations in this setup, we emphasise once more that this framework is indeed more general than the setup presented. In particular, the slowly varying functions $Q_0$ and $Q_1$ might arise from interactions with non-scalar field dark energy sources. For example, if the DE field's dynamics is negligible, our system reduces to DM coupled conformally to a cosmological constant. Another possibility motivated by string theory would be to have DM interacting with a 3-form DE field \cite{Koivisto_2013}, which could also reduce to this same phenomenology. 

\section{Perturbations} \label{sec:pert}

We now turn to the study of linear perturbations for our model. This will lead to the derivation of the sound speed, which is crucial to solving the fluid-scalar field equations of motion.

We follow the approach developed in Ref.~\cite{Hwang_2009} to compute the sound speed, considering only scalar perturbations. We adopt the following conventions for the metric scalar perturbations in a general gauge: 
\begin{align}
    \delta g_{00} & = -2  \Phi \mathcomma \\
    \delta g_{i0} &=a \nabla_i B \mathcomma \\
    \delta g_{ij} & =-2 a^2 (\delta_{ij}\Psi -\nabla_i\nabla_j E) \mathcomma
\end{align} 
where $\Phi$, $\Psi$, $B$ and $E$ are the four scalar degrees of freedom which can be expanded and decomposed in independently evolving Fourier modes $k$.
This results in the following equations of motion for the perturbations of the scalar fields $\delta \chi$ and $\delta \phi$:
\begin{eqnarray}
    \delta \ddot{\chi} &+& \left(3H + \frac{C_{,\phi}}{C}  \dot\phi \right)\delta \dot{\chi}+ \left(\frac{k^2}{a^2} + C(\phi) V_{,\chi\chi}\right)\delta\chi = \dot\chi \left(\dot\Phi+3 \dot{\Psi}-\frac{k}{a}B+\dot{E}\right) \nonumber \\
    &-& 2 C(\phi) V_{,\chi}\Phi 
    - C_{,\phi} V_{,\chi}\delta \phi 
    - \frac{C_{,\phi}}{C} \dot\chi \delta\dot\phi - \left(\frac{C_{,\phi}}{C}\right)_{,\phi} \dot\phi\dot\chi\delta\phi 
\end{eqnarray}
and 
\begin{equation} \label{eq:ddotphi}
    \delta \ddot{\phi}+3H\delta\dot{\phi}+\frac{k^2}{a^2}\delta \phi +m^2\delta \phi=\dot{\phi}\left(\dot{\Phi}+3\dot{\Psi}-\frac{k}{a}B +\dot{E}\right)+2(\ddot{\phi}+3H\dot{\phi})\Phi-\delta Q \mathcomma
\end{equation}
Where we have defined $\delta Q$ as the perturbation of the coupling function. As for the background evolution, we can expand it as
\begin{equation}
\delta Q= \delta Q_0+\phi \delta Q_1 + \delta \phi Q_1 \mathcomma
\end{equation}
such that Eq.~\eqref{eq:ddotphi} becomes
\begin{equation}\label{eq:kgpertfull}
        \delta \ddot{\phi}+3H\delta\dot{\phi}+\frac{k^2}{a^2}\delta \phi +m^2_{\rm eff}\delta \phi=\dot{\phi}\left(\dot{\Phi}+3\dot{\Psi}-\frac{k}{a}B +\dot{E}\right)+2(\ddot{\phi}+3H\dot{\phi})\Phi-\delta Q_0-\phi \delta Q_1 \mathperiod
\end{equation}
Note that the $\delta Q$ term encloses a term proportional to $\Phi$, arising due to the covariant derivatives in the definition of $Q$ in Eq.~\eqref{eq:coupling}. The exact form of $\delta Q_0 $ and $\delta Q_1$ for the particular setup given by Eqs.~\eqref{eq:vp}-\eqref{eq:couplingp} is the following: 
\begin{align}
\delta Q_0&=- \kappa \beta \left(2\dot{\chi}\delta \dot\chi -4 V_{,\chi}\delta \chi -2 \dot{\chi}^2 \Phi \right) \mathcomma \\
\delta Q_1& = 8 \kappa^2 \beta^2  V_{,\chi}\delta \chi  \mathperiod  
\end{align}
The perturbed perfect fluid quantities are defined in terms of the time-averaged oscillating DM scalar field and its perturbation:
\begin{equation}\label{eq:fluidrho}
    \delta \rho=\langle \dot{\phi} \delta\dot{\phi} -\dot{\phi}^2\Phi +m^2\phi \delta \phi \rangle \mathcomma
\end{equation}
\begin{equation}\label{eq:fluidp}
    \delta p=\langle \dot{\phi} \delta\dot{\phi} -\dot{\phi}^2\Phi -m^2\phi \delta \phi \rangle \mathcomma
\end{equation}
\begin{equation}\label{eq:fluidvel}
    \frac{a}{k}(\rho+p)(v-B)=\langle \dot\phi \delta \phi \rangle \mathcomma
\end{equation}
where it should be understood that the fluid variables are themselves averaged quantities. 
Bringing together Eqs.~\eqref{eq:kgpertfull}-\eqref{eq:fluidvel}, we arrive at the following expressions for the perturbed continuity equations:
\begin{equation}\label{eq:pertfluid1}
    \delta \dot{\rho}+3H(\delta \rho +\delta p )=(\rho+p)\left(3\dot{\Psi}+\dot{E}-\frac{k}{a} v\right) - \langle\delta Q \dot{\phi} \rangle-\langle Q \delta \dot{\phi}\rangle \mathcomma
\end{equation}
\begin{equation}\label{eq:pertfluid2}
    \frac{1}{a^4(\rho+p)} \frac{d}{dt}[a^4(\rho+p)(v-B)]
    =\frac{k}{a} \Phi+\frac{k}{a(\rho+p)}[\delta p-\langle Q\delta\phi \rangle] \mathperiod
\end{equation}

For convenience and to simplify the calculations, we consider the axion-comoving gauge, which amounts to setting $B=v$. Thus, in this context, Eq.~\eqref{eq:pertfluid2} reduces to 
\begin{equation}\label{eq:alphacomoving}
    \Phi = -\frac{\delta p}{\rho+p}+\frac{\langle Q\delta\phi \rangle}{\rho+p} \mathperiod
\end{equation}
Analogous to background fluid-field treatment, we specify the following ansatz for the perturbation of the $\phi$-field: 
\begin{equation}\label{eq:pertansatz}
   \delta \phi(k,t)=\delta \phi_+(k,t) \sin(m_{\rm eff}t)+\delta \phi_-(k,t) \cos(m_{\rm eff}t)+\delta{\mathcal{A}}(t) \mathperiod
\end{equation}
It is important to note that $\delta{\mathcal{A}}(t)$ is not just the perturbation of $A(t)$ defined in Eq.~\eqref{eq:at}, as will be shown explicitly below.

\begin{figure}[hh]
    \centering
    \includegraphics[height=0.6\linewidth]{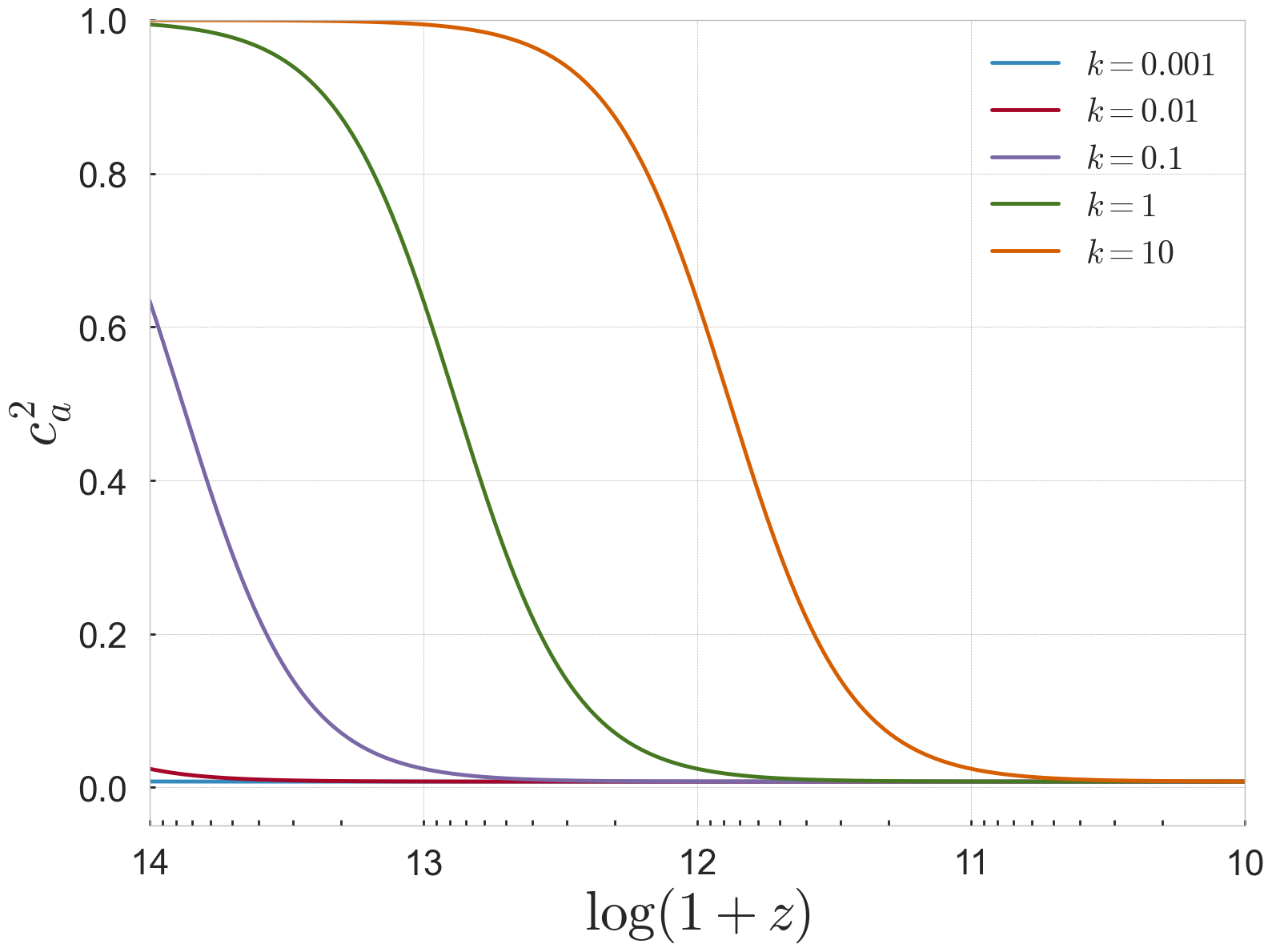}
    \caption{Evolution of the effective sound-speed in terms of redshift $z$, as defined in Eq.~\eqref{eq:cseff}, for a different set of $k$ values $\{0.001,0.01,0.1,1,10 \}$ (in Mpc$^{-1}$), for the following set of parameters: $m=10^{-17} \rm{eV}$, $\beta=5 \times 10^{14}$ and $\lambda=-0.2$.}
    \label{fig:c2k}
\end{figure}

\begin{figure}[hh]
    \centering
    \includegraphics[height=0.6\linewidth]{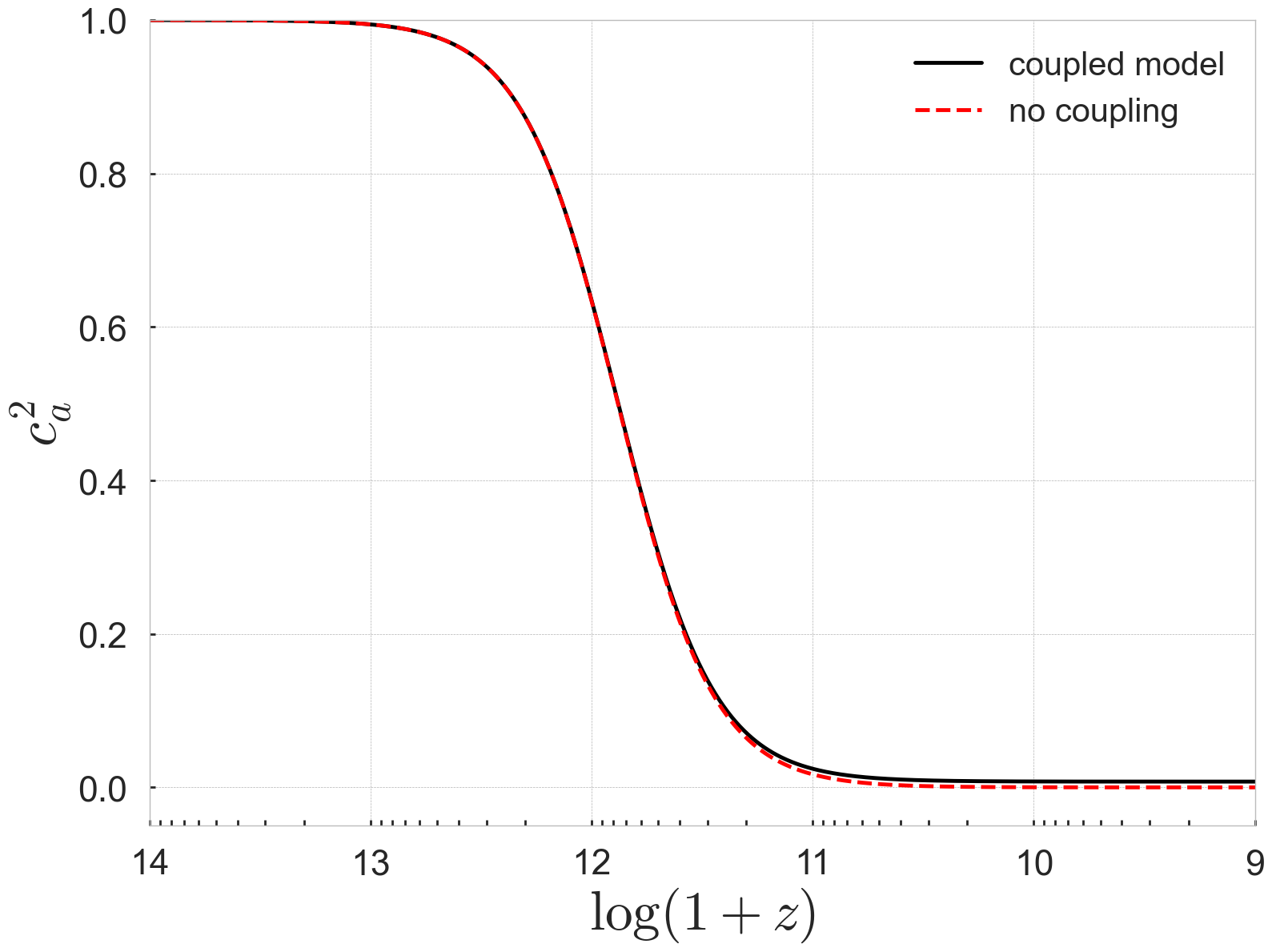}
    \caption{Evolution of the effective sound-speed for the coupled model, defined in Eq.~\eqref{eq:cseff}, compared with a non-coupled standard axion, with the same parameters used as in Fig.~\ref{fig:c2k} and $k $ = 10 Mpc$^{-1}$.}
    \label{fig:c2comparison}
\end{figure}

For the rest of the calculation, we will assume a quasi-static approximation following \cite{Hwang_2009}, and will also be discarding derivatives of $A$ and $\delta{\mathcal{A}}$ on similar grounds. In comoving gauge we have $\langle \dot{\phi} \delta \phi\rangle=0$ from Eq.~\eqref{eq:fluidvel}, which implies
\begin{equation}\label{eq:comovid1}
    \delta \phi_+\phi_-=\delta \phi_-\phi_+ \mathperiod
\end{equation}
Now we can solve Eq.~\eqref{eq:kgpertfull} to leading order in $H/m$, since the field is oscillating if $m \gg H$. This implies assuming that metric perturbations vary only on cosmological time scales $t \sim H^{-1} \gg m^{-1}$ and then grouping the terms in powers of $H/m$,
\begin{equation}\label{eq:kgpertHm}
    \delta \ddot\phi+\frac{k^2}{a^2}\delta \phi +m_{\rm eff}^2\delta \phi=-2(m^2_{\rm eff}\phi+Q_0)\Phi-\delta Q_0-\phi \delta Q_1 \mathperiod
\end{equation}
Once again, splitting the equation and matching the non-oscillating terms, we get:
\begin{equation}
    \delta\ddot{{\mathcal{A}}} +\frac{k^2}{a^2} \delta{\mathcal{A}} +m^2_{\rm eff} \delta{\mathcal{A}} =-\delta Q_0-\delta Q_1 A \mathcomma
\end{equation}
which, in the quasi-static approximation, yields 
\begin{equation}
    \delta{\mathcal{A}} = - \frac{\delta Q_0+ \delta Q_1 A}{\frac{k^2}{a^2}+m^2_{\rm eff}}= - \frac{\delta Q_0 - \frac{Q_0}{m_{\rm eff}^2}\delta Q_1 }{\frac{k^2}{a^2}+m^2_{\rm eff}} \mathperiod
\end{equation}
Substituting the ansatz in Eq.~\eqref{eq:pertansatz} into Eq.~\eqref{eq:kgpertHm} we obtain:
\begin{equation}\label{eq:alpha}
    \Phi=-\frac{1}{2} \left(\frac{a}{a_0}\right)^{3/2}
    \left(\frac{m_0}{m_{\rm eff}} \right)^{-1/2} \frac{\delta \phi_+}{\phi_+} \frac{k^2}{m^2_{\rm eff} a^2}-\frac{1}{2 m^2_{\rm eff}}\delta Q_1 \mathcomma
\end{equation}
up to leading order in $H/m$. Replacing this expression into the perturbed fluid equations for $\delta p$ and $\delta \rho$, Eqs.~\eqref{eq:fluidrho} and \eqref{eq:fluidp} we arrive at
\begin{eqnarray}
    \delta p &=&  \frac{a^{-\frac{3}{2}}}{2}\left(\frac{m_0}{m_{\rm eff}} \right)^{\frac{1}{2}} (\phi_+^2+\phi_-^2) \frac{\delta \phi_+}{\phi_+} m^2_{\rm eff} \left[ \frac{1}{2} \frac{k^2}{a^2 m^2_{\rm eff}}+\frac{Q_1}{m^2_{\rm eff}} \right]\nonumber \\
    &-& m^2 A \delta{\mathcal{A}} + \frac{a^{-3}}{4}\left(\frac{m_0}{m_{\rm eff}}\right)(\phi_+^2+\phi_-^2)\delta Q_1 \mathcomma \\
    \delta \rho &=&\frac{a^{-\frac{3}{2}}}{2}\left(\frac{m_0}{m_{\rm eff}} \right)^{\frac{1}{2}} (\phi_+^2+\phi_-^2) \frac{\delta \phi_+}{\phi_+} m^2_{\rm eff}\left[ \frac{1}{2} \frac{k^2}{a^2 m^2_{\rm eff}}+ 2 -\frac{Q_1}{m^2_{\rm eff}} \right]\nonumber \\
 &+& m^2 A \delta{\mathcal{A}} + \frac{a^{-3}}{4}\left(\frac{m_0}{m_{\rm eff}}\right)(\phi_+^2+\phi_-^2)\delta Q_1 \mathperiod
\end{eqnarray}
From here, and using Eq.~\eqref{eq:alpha}, we can compute the pressure perturbation, given by 
\begin{align}\label{eq:deltap}
    \delta p=c^2_{a}\delta \rho -m^2 A \delta{\mathcal{A}} (1+c^2_{a})+\frac{1}{2}\frac{\rho+p}{ m^2_{\rm eff}} \delta Q_1(1-c^2_{a}) \mathperiod
\end{align}
where we have defined the effective sound-speed 
\begin{equation} \label{eq:cseff}
    c^2_{a}=\frac{\frac{1}{2} \frac{k^2}{a^2 m^2_{\rm eff}}+\frac{Q_1}{m^2_{\rm eff}}}{\frac{1}{2} \frac{k^2}{a^2 m^2_{\rm eff}}+ 
 2-\frac{Q_1}{m^2_{\rm eff}}}= \frac{\frac{1}{2} \frac{k^2}{a^2 m^2}+\frac{Q_1}{m^2}}{\frac{1}{2} \frac{k^2}{a^2 m^2}+ 
 2+\frac{Q_1}{m^2}} \mathcomma
\end{equation}
and the second equality follows from the definition of $m^2_{\rm eff}$ in Eq.~\eqref{eq:meff}. It is important to note that this effective sound speed is akin to that of the non-interacting axion, presented in Refs.~\cite{Hwang_2009,Poulin_2018}. Accordingly, and as expected, Eq.~\eqref{eq:cseff} reduces to that case in the limit where the coupling $\beta$, and therefore $Q_1$, is zero. The effect of this coupling can be appreciated in Figs.~\ref{fig:c2k} and \ref{fig:c2comparison}. At early times, the $k^2/a^2m^2$ term dominates, and we get a similar behaviour to that of the non-interacting case. On the other hand, at late times, when $k^2/a^2m^2\ll Q_1$, the sound-speed becomes practically constant, according to $c^2_{a} \approx \frac{Q_1}{2m^2}$. As expected, in the absence of the coupling, $c^2_{a}$ falls to zero at small redshifts.

Furthermore, according to Eq.~\eqref{eq:deltap}, in contrast to the non-interacting case, the pressure perturbation is not precisely proportional to the density perturbation. This can be interpreted as a non-adiabatic contribution to the pressure perturbation caused by the interaction with the dark energy scalar field \cite{Christopherson_2009,Koshelev_2011,Carrillo_Gonz_lez_2018,Johnson:2020gzn}.

\section{Summary and Conclusions} \label{sec:conc}

In this study we have introduced a novel model of the dark sector, consisting of scalar field dark matter interacting with quintessence dark energy. Our approach involves a conformal coupling between DM and DE via a conformal coupling, with a unique twist: the conformal factor depends on the (fast oscillating) DM field rather than the slowly evolving DE field, as is typically the case. As a result, we have found an effective theory of interacting dark matter and dark energy in which the coupling term is linked to the energy density of DE rather than DM. Our framework not only offers a theoretical basis for many extensively studied models of interacting dark energy found in the existing literature, \textit{e.g.} in Refs.~\cite{Gavela:2009cy,Gavela:2010tm,Caldera-Cabral:2009hoy,DiValentino:2017iww,DiValentino:2019jae,Zhao:2022ycr,Zhai:2023yny,Giare:2024ytc}, but also introduces significant differences. 
Namely, within our setup, the coupling between DM and DE serves to offset the oscillations of the scalar field, resulting in a non-zero average for its equation of state parameter. Upon averaging over these rapid oscillations, we have derived a fluid-field description of the system, revealing a non-zero average value for the field determined by the coupling.
Consequently, the effective DM fluid exhibits a non-zero physical pressure, starkly contrasting with standard IDE models. This physical pressure transitions from slightly positive at early times to negative after DM-DE equality. The main observational effect most likely comes from the change in the DM equation of state at very late times, at a redshift smaller than the redshift of the DE-DM equality. Thus, our work carries implications in the context of searches for non-standard CDM physics and provides theoretical support for such models (see \cite{Ilic:2020onu,Pan:2022qrr,khurshudyan2024constraints}). Notably, the non-standard equation of state of DM impacts the analysis and interpretation of cosmological observational data, such as Pantheon+ \cite{Brout:2022vxf} or the recently published DESI data \cite{DESI:2024mwx}.

At the linear perturbation level, we have computed the pressure perturbation and sound speed of the averaged DM fluid. Our findings reveal terms proportional to the density perturbation $\delta \rho$ plus non-adiabatic pressure terms depending on the perturbation of the DE scalar field. Notably, due to the coupling, the adiabatic term deviates from the standard axion scalar field scenario. This results in an effective sound speed that remains non-zero across all scales, unlike the uncoupled case where it vanishes at small $k$. Consequently, we anticipate a slight power suppression in the matter power spectrum at all scales within the interacting scenario considered here.
Furthermore, the influence of non-adiabatic pressure contributions on observables in this context warrants further investigation. A comprehensive study of these points requires a full implementation of the theory in publicly available Boltzmann codes such as \texttt{CLASS} or \texttt{CAMB}\footnote{\href{https://github.com/cmbant/CAMB}{https://github.com/cmbant/CAMB}} \cite{Lewis:1999bs}, a task left for future work.

\acknowledgments We are grateful to Eleonora Di Valentino, William Giarè and Vivian Poulin for useful discussions and encouragement. EMT is supported by funding from the European Research Council (ERC) under the European Union's HORIZON-ERC-2022 (grant agreement no. 101076865). EMT was also supported by the grant SFRH/BD/143231/2019 from Funda\c{c}\~ao para a Ci\^encia e a Tecnologia (FCT) during part of this work. CvdB is supported (in part) by the Lancaster–Sheffield Consortium for Fundamental Physics under STFC grant: ST/X000621/1. NJN acknowledges support from FCT via the following projects: 
UIDB/04434/2020 \& UIDP/04434/2020 and PTDC/FIS-AST/0054/2021. This article is based upon work from COST Action CA21136 Addressing observational tensions in cosmology with systematics and fundamental physics (Cosmoverse) supported by COST (European Cooperation in Science and Technology).

\section*{Appendix A}\label{AppA}

In the main body of the paper, we considered a conformal coupling function $C$, which depends only on the DM field value, \textit{i.e.} we considered the case of $C(\phi)$. It is possible to consider a more general scenario, in which the $C$ depends both on the field and its kinetic term $X = -\frac{1}{2}g^{\mu\nu}\partial_\mu \phi \partial_\nu \phi$. For completeness, in this appendix, we provide the equations for that case, that is, $C \equiv C(\phi, X)$. The equation of motion for the dark energy field $\chi$ can be derived from the action in Eq.~\eqref{eq:new_coupledquintessence_action} and reads
\begin{equation}\label{eq:KGequation}
\nabla^\mu \nabla_\mu  \chi - C \frac{dV}{d\chi} = - \frac{\nabla_\alpha C}{C}\nabla^\alpha \chi \mathperiod
\end{equation}
In the same way, the equation for the DM field $\phi$ is 
\begin{equation}
\nabla^\mu \nabla_\mu \phi - \frac{\partial U}{\partial \phi} = Q \mathcomma
\end{equation}
where 
\begin{eqnarray}\label{eq:coupling_full}
Q &=& \nabla_\mu\left( \frac{\partial L_\chi}{\partial(\nabla_\mu \phi)} \right) - \frac{\partial L_\chi}{\partial \phi} \nonumber \\
&=& \nabla_\mu \left[\nabla^\mu\phi \left(Cg^{\alpha\beta}\nabla_\alpha\chi \nabla_\beta \chi + 4C^2U \right)\frac{C_{,X}}{C}\right] + \frac{C_{,\phi}}{2C}\left[ Cg^{\alpha\beta}\nabla_\alpha\chi \nabla_\beta \chi + 4C^2U  \right] \mathperiod
\end{eqnarray}
The energy momentum tensor for $\phi$ and $\chi$ are given by 
\begin{eqnarray}
T_{\mu\nu}^{(\phi)} = \nabla_\mu \phi \nabla_\nu \phi - g_{\mu\nu} \left( \frac{1}{2}g^{\alpha\beta} \nabla_\alpha\phi\nabla_\beta \phi + V(\phi) \right) \mathcomma
\end{eqnarray}
\begin{eqnarray}
T_{\mu\nu}^{(\chi)} &=& C \nabla_\mu \chi \nabla_\nu \chi - g_{\mu\nu} \left( \frac{C}{2}g^{\alpha\beta} \nabla_\alpha\chi\nabla_\beta \chi + C^2 U(\chi) \right) \nonumber \\
&-& \frac{C_{,X}}{2C}\nabla_\mu \phi \nabla_\nu \phi \left[ Cg^{\alpha\beta}\nabla_\alpha\chi \nabla_\beta \chi + 4C^2U   \right] \mathperiod
\end{eqnarray}
We have the following conservation equations:
\begin{equation}
   \nabla^\mu T_{\mu\nu}^{(\phi)} = Q \nabla_\nu \phi; ~~~\nabla^\mu T_{\mu\nu}^{(\chi)} = -Q \nabla_\nu \phi \mathperiod
\end{equation}
Note that 
\begin{equation}
T^{\mu}_{(\chi)\mu} = - \left[C\nabla^\mu \chi \nabla_\mu \chi + 4C^2 U(\chi) \right]\left( 1 + \frac{C_{,X}}{2C} \nabla^\mu \phi \nabla_\mu \phi  \right)\mathcomma   
\end{equation}
and so the two terms in Eq.~\eqref{eq:coupling_full} are proportional to $T^{\mu}_{(\chi)\mu}$. 

In full generality, these equations are difficult to analyse and to recast into a fluid-field description. Nevertheless, the sub-case of $C\equiv C(X)$ might be worth investigating. We leave this for future work. 

\section*{Appendix B}\label{AppB}
In the main body of the paper we have dealt with an interaction mediated by a conformal coupling which depends on the fast oscillating DM field. In the following we briefly present another example in which DE and DM couple, but not via a conformal coupling. The phenomenology turns out to be very similar to the model presented in Sections \ref{sec:fluid} and \ref{sec:pert}. 

The action we consider is 
 \begin{eqnarray}\label{eq:alternative_action}
{\cal S} &=& \int d^4 x \sqrt{-g}\left( \frac{M_{\rm Pl}^2}{2} {\cal R} - \frac{1}{2}g^{\mu\nu} \partial_\mu \phi \partial_\nu\phi - U(\phi) - \frac{1}{2}g^{\mu\nu} \partial_\mu \chi \partial_\nu\chi - V(\chi) \right) \\ \nonumber 
&-& \int d^4 x \sqrt{-g}\left(\phi Q_0\left(\chi, X^{(\chi)}\right) + \frac{1}{2}\phi^2 Q_1 \left(\chi , X^{(\chi)} \right)\right) + 
{\cal S}_{\rm SM} \mathcomma
\end{eqnarray}
where $Q_0$ and $Q_1$ are coupling functions, depending of the slowly varying DE field $\chi$ and its kinetic term $ X^{(\chi)} = - \frac{1}{2}g^{\mu\nu} \partial_\mu \chi \partial_\nu\chi$. Let us focus here on the case in which $Q_0$ and $Q_1$ depend on $\chi$ only. Then, in a flat FRW universe, the equation of motion for $\phi$ is given by (using $U = \frac{1}{2} m^2 \phi^2$)
\begin{equation}
    \ddot{\phi} + 3H \dot{\phi} +  m^2\phi = -Q(t) \mathperiod
\end{equation}
where $Q(t) = Q_0(\chi) + Q_1(\chi)\phi$. This equation is of the same form as Eq.~\eqref{eq:kgeqQ}. Assuming that $\chi$ evolves slowly as it represents DE, an analysis similar to that performed in Section \ref{sec:fluid} would apply for this theory, with the displacement $A$ in \eqref{eq:at} now given by 
\begin{eqnarray} 
    A(t) \approx -\frac{Q_0(\chi)}{m^2_{\rm eff}} \mathcomma
\end{eqnarray}
and with an effective mass that is still given by Eq. \eqref{eq:meff}. 

The equation for the DE field reads
\begin{equation}
    \ddot \chi + 3H\dot\chi + V_{,\chi} + \phi Q_{0,\chi} + \frac{1}{2}\phi^2 Q_{1,\chi} = 0 \mathperiod
\end{equation}
We can now employ the time-averaged behaviour of the DM field. The end result is that the dynamics of the DE field is described by 
\begin{equation}
    \ddot \chi + 3H\dot\chi + V_{,\chi} - \frac{Q_0}{m_{\rm eff}^2}Q_{0,\chi} + \frac{2\langle \rho_{\phi} \rangle + \frac{Q_0^2}{m_{\rm eff}^2}}{2 \left(m_{\rm eff}^2 + m^2 \right)}Q_{1,\chi}= 0 \mathcomma 
\end{equation}
where we have used Eqs.~\eqref{eq:phi2variance} and \eqref{avenergy}. 

At the linear perturbations level, the equation for the DM field is given by Eq.~\eqref{eq:ddotphi} and so the expression for the sound-speed is unchanged.

\bibliographystyle{JHEP}
\bibliography{biblio.bib}

\providecommand{\href}[2]{#2}\begingroup\raggedright\begin{thebibliography}{10}

\bibitem{Peebles:2022bya}
P.J.E.~Peebles, \emph{{Cosmology\textquoteright{}s Century: An Inside History of Our Modern Understanding of the Universe}}, Princeton University Press (4, 2022).

\bibitem{Abdalla:2022yfr}
E.~Abdalla et~al., \emph{{Cosmology intertwined: A review of the particle physics, astrophysics, and cosmology associated with the cosmological tensions and anomalies}}, \href{https://doi.org/10.1016/j.jheap.2022.04.002}{\emph{JHEAp} {\bfseries 34} (2022) 49} [\href{https://arxiv.org/abs/2203.06142}{{\ttfamily 2203.06142}}].

\bibitem{1980ApJ...238..471R}
V.C.~{Rubin}, J.~{Ford}, W.~K. and N.~{Thonnard}, \emph{{Rotational properties of 21 SC galaxies with a large range of luminosities and radii, from NGC 4605 (R=4kpc) to UGC 2885 (R=122kpc).}}, \href{https://doi.org/10.1086/158003}{\emph{Astrophys. J.} {\bfseries 238} (1980) 471}.

\bibitem{1933AcHPh...6..110Z}
F.~{Zwicky}, \emph{{Die Rotverschiebung von extragalaktischen Nebeln}}, {\emph{Helvetica Physica Acta} {\bfseries 6} (1933) 110}.

\bibitem{1937ApJ....86..217Z}
F.~{Zwicky}, \emph{{On the Masses of Nebulae and of Clusters of Nebulae}}, \href{https://doi.org/10.1086/143864}{\emph{Astrophys. J.} {\bfseries 86} (1937) 217}.

\bibitem{Clowe:2006eq}
D.~Clowe, M.~Bradac, A.H.~Gonzalez, M.~Markevitch, S.W.~Randall, C.~Jones et~al., \emph{{A direct empirical proof of the existence of dark matter}}, \href{https://doi.org/10.1086/508162}{\emph{Astrophys. J. Lett.} {\bfseries 648} (2006) L109} [\href{https://arxiv.org/abs/astro-ph/0608407}{{\ttfamily astro-ph/0608407}}].

\bibitem{Planck:2018vyg}
{\scshape Planck} collaboration, \emph{{Planck 2018 results. VI. Cosmological parameters}}, \href{https://doi.org/10.1051/0004-6361/201833910}{\emph{Astron. Astrophys.} {\bfseries 641} (2020) A6} [\href{https://arxiv.org/abs/1807.06209}{{\ttfamily 1807.06209}}].

\bibitem{2013ApJS..208...19H}
G.~{Hinshaw}, D.~{Larson}, E.~{Komatsu}, D.N.~{Spergel}, C.L.~{Bennett}, J.~{Dunkley} et~al., \emph{{Nine-year Wilkinson Microwave Anisotropy Probe (WMAP) Observations: Cosmological Parameter Results}}, \href{https://doi.org/10.1088/0067-0049/208/2/19}{\emph{Astrophys. J. Sup.} {\bfseries 208} (2013) 19} [\href{https://arxiv.org/abs/1212.5226}{{\ttfamily 1212.5226}}].

\bibitem{Gottel:2024cfj}
A.S.~G\"ottel, A.~Ejlli, K.~Karan, S.M.~Vermeulen, L.~Aiello, V.~Raymond et~al., \emph{{Searching for scalar field dark matter with LIGO}},  \href{https://arxiv.org/abs/2401.18076}{{\ttfamily 2401.18076}}.

\bibitem{XENON:2018voc}
{\scshape XENON} collaboration, \emph{{Dark Matter Search Results from a One Ton-Year Exposure of XENON1T}}, \href{https://doi.org/10.1103/PhysRevLett.121.111302}{\emph{Phys. Rev. Lett.} {\bfseries 121} (2018) 111302} [\href{https://arxiv.org/abs/1805.12562}{{\ttfamily 1805.12562}}].

\bibitem{LUX:2016ggv}
{\scshape LUX} collaboration, \emph{{Results from a search for dark matter in the complete LUX exposure}}, \href{https://doi.org/10.1103/PhysRevLett.118.021303}{\emph{Phys. Rev. Lett.} {\bfseries 118} (2017) 021303} [\href{https://arxiv.org/abs/1608.07648}{{\ttfamily 1608.07648}}].

\bibitem{Kahlhoefer:2017dnp}
F.~Kahlhoefer, \emph{{Review of LHC Dark Matter Searches}}, \href{https://doi.org/10.1142/S0217751X1730006X}{\emph{Int. J. Mod. Phys. A} {\bfseries 32} (2017) 1730006} [\href{https://arxiv.org/abs/1702.02430}{{\ttfamily 1702.02430}}].

\bibitem{Bertone:2018krk}
G.~Bertone and T.~Tait, M.~P., \emph{{A new era in the search for dark matter}}, \href{https://doi.org/10.1038/s41586-018-0542-z}{\emph{Nature} {\bfseries 562} (2018) 51} [\href{https://arxiv.org/abs/1810.01668}{{\ttfamily 1810.01668}}].

\bibitem{Graham:2015ouw}
P.W.~Graham, I.G.~Irastorza, S.K.~Lamoreaux, A.~Lindner and K.A.~van Bibber, \emph{{Experimental Searches for the Axion and Axion-Like Particles}}, \href{https://doi.org/10.1146/annurev-nucl-102014-022120}{\emph{Ann. Rev. Nucl. Part. Sci.} {\bfseries 65} (2015) 485} [\href{https://arxiv.org/abs/1602.00039}{{\ttfamily 1602.00039}}].

\bibitem{Choi:2020rgn}
K.~Choi, S.H.~Im and C.~Sub~Shin, \emph{{Recent Progress in the Physics of Axions and Axion-Like Particles}}, \href{https://doi.org/10.1146/annurev-nucl-120720-031147}{\emph{Ann. Rev. Nucl. Part. Sci.} {\bfseries 71} (2021) 225} [\href{https://arxiv.org/abs/2012.05029}{{\ttfamily 2012.05029}}].

\bibitem{Amruth:2023xqj}
A.~Amruth et~al., \emph{{Einstein rings modulated by wavelike dark matter from anomalies in gravitationally lensed images}}, \href{https://doi.org/10.1038/s41550-023-01943-9}{\emph{Nature Astron.} {\bfseries 7} (2023) 736} [\href{https://arxiv.org/abs/2304.09895}{{\ttfamily 2304.09895}}].

\bibitem{Hu:2000ke}
W.~Hu, R.~Barkana and A.~Gruzinov, \emph{{Cold and fuzzy dark matter}}, \href{https://doi.org/10.1103/PhysRevLett.85.1158}{\emph{Phys. Rev. Lett.} {\bfseries 85} (2000) 1158} [\href{https://arxiv.org/abs/astro-ph/0003365}{{\ttfamily astro-ph/0003365}}].

\bibitem{Dentler:2021zij}
M.~Dentler, D.J.E.~Marsh, R.~Hlo\v{z}ek, A.~Lagu\"e, K.K.~Rogers and D.~Grin, \emph{{Fuzzy dark matter and the Dark Energy Survey Year 1 data}}, \href{https://doi.org/10.1093/mnras/stac1946}{\emph{Mon. Not. Roy. Astron. Soc.} {\bfseries 515} (2022) 5646} [\href{https://arxiv.org/abs/2111.01199}{{\ttfamily 2111.01199}}].

\bibitem{Svrcek:2006yi}
P.~Svrcek and E.~Witten, \emph{{Axions In String Theory}}, \href{https://doi.org/10.1088/1126-6708/2006/06/051}{\emph{JHEP} {\bfseries 06} (2006) 051} [\href{https://arxiv.org/abs/hep-th/0605206}{{\ttfamily hep-th/0605206}}].

\bibitem{Arvanitaki:2009fg}
A.~Arvanitaki, S.~Dimopoulos, S.~Dubovsky, N.~Kaloper and J.~March-Russell, \emph{{String Axiverse}}, \href{https://doi.org/10.1103/PhysRevD.81.123530}{\emph{Phys. Rev. D} {\bfseries 81} (2010) 123530} [\href{https://arxiv.org/abs/0905.4720}{{\ttfamily 0905.4720}}].

\bibitem{Hlozek:2014lca}
R.~Hlozek, D.~Grin, D.J.E.~Marsh and P.G.~Ferreira, \emph{{A search for ultralight axions using precision cosmological data}}, \href{https://doi.org/10.1103/PhysRevD.91.103512}{\emph{Phys. Rev. D} {\bfseries 91} (2015) 103512} [\href{https://arxiv.org/abs/1410.2896}{{\ttfamily 1410.2896}}].

\bibitem{Bullock:2017xww}
J.S.~Bullock and M.~Boylan-Kolchin, \emph{{Small-Scale Challenges to the $\Lambda$CDM Paradigm}}, \href{https://doi.org/10.1146/annurev-astro-091916-055313}{\emph{Ann. Rev. Astron. Astrophys.} {\bfseries 55} (2017) 343} [\href{https://arxiv.org/abs/1707.04256}{{\ttfamily 1707.04256}}].

\bibitem{Weinberg:2013aya}
D.H.~Weinberg, J.S.~Bullock, F.~Governato, R.~Kuzio~de Naray and A.H.G.~Peter, \emph{{Cold dark matter: controversies on small scales}}, \href{https://doi.org/10.1073/pnas.1308716112}{\emph{Proc. Nat. Acad. Sci.} {\bfseries 112} (2015) 12249} [\href{https://arxiv.org/abs/1306.0913}{{\ttfamily 1306.0913}}].

\bibitem{Hui:2016ltb}
L.~Hui, J.P.~Ostriker, S.~Tremaine and E.~Witten, \emph{{Ultralight scalars as cosmological dark matter}}, \href{https://doi.org/10.1103/PhysRevD.95.043541}{\emph{Phys. Rev. D} {\bfseries 95} (2017) 043541} [\href{https://arxiv.org/abs/1610.08297}{{\ttfamily 1610.08297}}].

\bibitem{Rogers:2023ezo}
K.K.~Rogers, R.~Hlo\v{z}ek, A.~Lagu\"e, M.M.~Ivanov, O.H.E.~Philcox, G.~Cabass et~al., \emph{{Ultra-light axions and the S $_{8}$ tension: joint constraints from the cosmic microwave background and galaxy clustering}}, \href{https://doi.org/10.1088/1475-7516/2023/06/023}{\emph{JCAP} {\bfseries 06} (2023) 023} [\href{https://arxiv.org/abs/2301.08361}{{\ttfamily 2301.08361}}].

\bibitem{Lague:2020htq}
A.~Lagu\"e, J.R.~Bond, R.~Hlo\v{z}ek, D.J.E.~Marsh and L.~S\"oding, \emph{{Evolving ultralight scalars into non-linearity with Lagrangian perturbation theory}}, \href{https://doi.org/10.1093/mnras/stab601}{\emph{Mon. Not. Roy. Astron. Soc.} {\bfseries 504} (2021) 2391} [\href{https://arxiv.org/abs/2004.08482}{{\ttfamily 2004.08482}}].

\bibitem{Green:1987sp}
M.B.~Green, J.H.~Schwarz and E.~Witten, \emph{{SUPERSTRING THEORY. VOL. 1: INTRODUCTION}}, Cambridge Monographs on Mathematical Physics (7, 1988).

\bibitem{Amendola:2005ad}
L.~Amendola and R.~Barbieri, \emph{{Dark matter from an ultra-light pseudo-Goldsone-boson}}, \href{https://doi.org/10.1016/j.physletb.2006.08.069}{\emph{Phys. Lett. B} {\bfseries 642} (2006) 192} [\href{https://arxiv.org/abs/hep-ph/0509257}{{\ttfamily hep-ph/0509257}}].

\bibitem{Hlozek:2016lzm}
R.~Hlo\v{z}ek, D.J.E.~Marsh, D.~Grin, R.~Allison, J.~Dunkley and E.~Calabrese, \emph{{Future CMB tests of dark matter: Ultralight axions and massive neutrinos}}, \href{https://doi.org/10.1103/PhysRevD.95.123511}{\emph{Phys. Rev. D} {\bfseries 95} (2017) 123511} [\href{https://arxiv.org/abs/1607.08208}{{\ttfamily 1607.08208}}].

\bibitem{Hlozek:2017zzf}
R.~Hlozek, D.J.E.~Marsh and D.~Grin, \emph{{Using the Full Power of the Cosmic Microwave Background to Probe Axion Dark Matter}}, \href{https://doi.org/10.1093/mnras/sty271}{\emph{Mon. Not. Roy. Astron. Soc.} {\bfseries 476} (2018) 3063} [\href{https://arxiv.org/abs/1708.05681}{{\ttfamily 1708.05681}}].

\bibitem{Farren:2021jcd}
G.S.~Farren, D.~Grin, A.H.~Jaffe, R.~Hlo\v{z}ek and D.J.E.~Marsh, \emph{{Ultralight axions and the kinetic Sunyaev-Zel\textquoteright{}dovich effect}}, \href{https://doi.org/10.1103/PhysRevD.105.063513}{\emph{Phys. Rev. D} {\bfseries 105} (2022) 063513} [\href{https://arxiv.org/abs/2109.13268}{{\ttfamily 2109.13268}}].

\bibitem{Irsic:2017yje}
V.~Ir\v{s}i\v{c}, M.~Viel, M.G.~Haehnelt, J.S.~Bolton and G.D.~Becker, \emph{{First constraints on fuzzy dark matter from Lyman-$\alpha$ forest data and hydrodynamical simulations}}, \href{https://doi.org/10.1103/PhysRevLett.119.031302}{\emph{Phys. Rev. Lett.} {\bfseries 119} (2017) 031302} [\href{https://arxiv.org/abs/1703.04683}{{\ttfamily 1703.04683}}].

\bibitem{Armengaud:2017nkf}
E.~Armengaud, N.~Palanque-Delabrouille, C.~Y\`eche, D.J.E.~Marsh and J.~Baur, \emph{{Constraining the mass of light bosonic dark matter using SDSS Lyman-$\alpha$ forest}}, \href{https://doi.org/10.1093/mnras/stx1870}{\emph{Mon. Not. Roy. Astron. Soc.} {\bfseries 471} (2017) 4606} [\href{https://arxiv.org/abs/1703.09126}{{\ttfamily 1703.09126}}].

\bibitem{Kobayashi:2017jcf}
T.~Kobayashi, R.~Murgia, A.~De~Simone, V.~Ir\v{s}i\v{c} and M.~Viel, \emph{{Lyman-$\alpha$ constraints on ultralight scalar dark matter: Implications for the early and late universe}}, \href{https://doi.org/10.1103/PhysRevD.96.123514}{\emph{Phys. Rev. D} {\bfseries 96} (2017) 123514} [\href{https://arxiv.org/abs/1708.00015}{{\ttfamily 1708.00015}}].

\bibitem{Nori:2018pka}
M.~Nori, R.~Murgia, V.~Ir\v{s}i\v{c}, M.~Baldi and M.~Viel, \emph{{Lyman $\alpha$ forest and non-linear structure characterization in Fuzzy Dark Matter cosmologies}}, \href{https://doi.org/10.1093/mnras/sty2888}{\emph{Mon. Not. Roy. Astron. Soc.} {\bfseries 482} (2019) 3227} [\href{https://arxiv.org/abs/1809.09619}{{\ttfamily 1809.09619}}].

\bibitem{Rogers:2020ltq}
K.K.~Rogers and H.V.~Peiris, \emph{{Strong Bound on Canonical Ultralight Axion Dark Matter from the Lyman-Alpha Forest}}, \href{https://doi.org/10.1103/PhysRevLett.126.071302}{\emph{Phys. Rev. Lett.} {\bfseries 126} (2021) 071302} [\href{https://arxiv.org/abs/2007.12705}{{\ttfamily 2007.12705}}].

\bibitem{Hayashi:2021xxu}
K.~Hayashi, E.G.M.~Ferreira and H.Y.J.~Chan, \emph{{Narrowing the Mass Range of Fuzzy Dark Matter with Ultrafaint Dwarfs}}, \href{https://doi.org/10.3847/2041-8213/abf501}{\emph{Astrophys. J. Lett.} {\bfseries 912} (2021) L3} [\href{https://arxiv.org/abs/2102.05300}{{\ttfamily 2102.05300}}].

\bibitem{Dalal:2022rmp}
N.~Dalal and A.~Kravtsov, \emph{{Excluding fuzzy dark matter with sizes and stellar kinematics of ultrafaint dwarf galaxies}}, \href{https://doi.org/10.1103/PhysRevD.106.063517}{\emph{Phys. Rev. D} {\bfseries 106} (2022) 063517} [\href{https://arxiv.org/abs/2203.05750}{{\ttfamily 2203.05750}}].

\bibitem{Goldstein:2022pxu}
I.S.~Goldstein, S.M.~Koushiappas and M.G.~Walker, \emph{{Viability of ultralight bosonic dark matter in dwarf galaxies}}, \href{https://doi.org/10.1103/PhysRevD.106.063010}{\emph{Phys. Rev. D} {\bfseries 106} (2022) 063010} [\href{https://arxiv.org/abs/2206.05244}{{\ttfamily 2206.05244}}].

\bibitem{Lague:2021frh}
A.~Lagu\"e, J.R.~Bond, R.~Hlo\v{z}ek, K.K.~Rogers, D.J.E.~Marsh and D.~Grin, \emph{{Constraining ultralight axions with galaxy surveys}}, \href{https://doi.org/10.1088/1475-7516/2022/01/049}{\emph{JCAP} {\bfseries 01} (2022) 049} [\href{https://arxiv.org/abs/2104.07802}{{\ttfamily 2104.07802}}].

\bibitem{Kunkel:2022ldl}
A.~Kunkel, T.~Chiueh and B.M.~Sch\"afer, \emph{{A weak lensing perspective on non-linear structure formation with fuzzy dark matter}}, \href{https://doi.org/10.1093/mnras/stad3737}{\emph{Mon. Not. Roy. Astron. Soc.} {\bfseries 527} (2023) 10538} [\href{https://arxiv.org/abs/2211.01523}{{\ttfamily 2211.01523}}].

\bibitem{Hotinli:2021vxg}
S.C.~Hotinli, D.J.E.~Marsh and M.~Kamionkowski, \emph{{Probing ultralight axions with the 21-cm signal during cosmic dawn}}, \href{https://doi.org/10.1103/PhysRevD.106.043529}{\emph{Phys. Rev. D} {\bfseries 106} (2022) 043529} [\href{https://arxiv.org/abs/2112.06943}{{\ttfamily 2112.06943}}].

\bibitem{Bauer:2020zsj}
J.B.~Bauer, D.J.E.~Marsh, R.~Hlo\v{z}ek, H.~Padmanabhan and A.~Lagu\"e, \emph{{Intensity Mapping as a Probe of Axion Dark Matter}}, \href{https://doi.org/10.1093/mnras/staa3300}{\emph{Mon. Not. Roy. Astron. Soc.} {\bfseries 500} (2020) 3162} [\href{https://arxiv.org/abs/2003.09655}{{\ttfamily 2003.09655}}].

\bibitem{Flitter:2022pzf}
J.~Flitter and E.D.~Kovetz, \emph{{Closing the window on fuzzy dark matter with the 21-cm signal}}, \href{https://doi.org/10.1103/PhysRevD.106.063504}{\emph{Phys. Rev. D} {\bfseries 106} (2022) 063504} [\href{https://arxiv.org/abs/2207.05083}{{\ttfamily 2207.05083}}].

\bibitem{EPTA:2023xxk}
{\scshape EPTA} collaboration, \emph{{The second data release from the European Pulsar Timing Array: V. Implications for massive black holes, dark matter and the early Universe}},  \href{https://arxiv.org/abs/2306.16227}{{\ttfamily 2306.16227}}.

\bibitem{NANOGrav:2023gor}
{\scshape NANOGrav} collaboration, \emph{{The NANOGrav 15 yr Data Set: Evidence for a Gravitational-wave Background}}, \href{https://doi.org/10.3847/2041-8213/acdac6}{\emph{Astrophys. J. Lett.} {\bfseries 951} (2023) L8} [\href{https://arxiv.org/abs/2306.16213}{{\ttfamily 2306.16213}}].

\bibitem{Khmelnitsky:2013lxt}
A.~Khmelnitsky and V.~Rubakov, \emph{{Pulsar timing signal from ultralight scalar dark matter}}, \href{https://doi.org/10.1088/1475-7516/2014/02/019}{\emph{JCAP} {\bfseries 02} (2014) 019} [\href{https://arxiv.org/abs/1309.5888}{{\ttfamily 1309.5888}}].

\bibitem{Porayko:2014rfa}
N.K.~Porayko and K.A.~Postnov, \emph{{Constraints on ultralight scalar dark matter from pulsar timing}}, \href{https://doi.org/10.1103/PhysRevD.90.062008}{\emph{Phys. Rev. D} {\bfseries 90} (2014) 062008} [\href{https://arxiv.org/abs/1408.4670}{{\ttfamily 1408.4670}}].

\bibitem{Porayko:2018sfa}
N.K.~Porayko et~al., \emph{{Parkes Pulsar Timing Array constraints on ultralight scalar-field dark matter}}, \href{https://doi.org/10.1103/PhysRevD.98.102002}{\emph{Phys. Rev. D} {\bfseries 98} (2018) 102002} [\href{https://arxiv.org/abs/1810.03227}{{\ttfamily 1810.03227}}].

\bibitem{Xia:2023hov}
Z.-Q.~Xia, T.-P.~Tang, X.~Huang, Q.~Yuan and Y.-Z.~Fan, \emph{{Constraining ultralight dark matter using the Fermi-LAT pulsar timing array}}, \href{https://doi.org/10.1103/PhysRevD.107.L121302}{\emph{Phys. Rev. D} {\bfseries 107} (2023) L121302} [\href{https://arxiv.org/abs/2303.17545}{{\ttfamily 2303.17545}}].

\bibitem{EuropeanPulsarTimingArray:2023egv}
{\scshape European Pulsar Timing Array} collaboration, \emph{{Second Data Release from the European Pulsar Timing Array: Challenging the Ultralight Dark Matter Paradigm}}, \href{https://doi.org/10.1103/PhysRevLett.131.171001}{\emph{Phys. Rev. Lett.} {\bfseries 131} (2023) 171001} [\href{https://arxiv.org/abs/2306.16228}{{\ttfamily 2306.16228}}].

\bibitem{Wetterich:1987fm}
C.~Wetterich, \emph{{Cosmology and the Fate of Dilatation Symmetry}}, \href{https://doi.org/10.1016/0550-3213(88)90193-9}{\emph{Nucl. Phys. B} {\bfseries 302} (1988) 668} [\href{https://arxiv.org/abs/1711.03844}{{\ttfamily 1711.03844}}].

\bibitem{Peebles:1987ek}
P.J.E.~Peebles and B.~Ratra, \emph{{Cosmology with a Time Variable Cosmological Constant}}, \href{https://doi.org/10.1086/185100}{\emph{Astrophys. J. Lett.} {\bfseries 325} (1988) L17}.

\bibitem{Ratra:1987rm}
B.~Ratra and P.J.E.~Peebles, \emph{{Cosmological Consequences of a Rolling Homogeneous Scalar Field}}, \href{https://doi.org/10.1103/PhysRevD.37.3406}{\emph{Phys. Rev. D} {\bfseries 37} (1988) 3406}.

\bibitem{Wetterich:1994bg}
C.~Wetterich, \emph{{The Cosmon model for an asymptotically vanishing time dependent cosmological 'constant'}}, {\emph{Astron. Astrophys.} {\bfseries 301} (1995) 321} [\href{https://arxiv.org/abs/hep-th/9408025}{{\ttfamily hep-th/9408025}}].

\bibitem{Caldwell:1997ii}
R.R.~Caldwell, R.~Dave and P.J.~Steinhardt, \emph{{Cosmological imprint of an energy component with general equation of state}}, \href{https://doi.org/10.1103/PhysRevLett.80.1582}{\emph{Phys. Rev. Lett.} {\bfseries 80} (1998) 1582} [\href{https://arxiv.org/abs/astro-ph/9708069}{{\ttfamily astro-ph/9708069}}].

\bibitem{Koivisto_2013}
T.S.~Koivisto and N.J.~Nunes, \emph{Coupled three-form dark energy}, \href{https://doi.org/10.1103/physrevd.88.123512}{\emph{Physical Review D} {\bfseries 88} (2013) }.

\bibitem{Amendola:2015ksp}
L.~Amendola and S.~Tsujikawa, \emph{{Dark Energy}: {Theory and Observations}}, Cambridge University Press (1, 2015).

\bibitem{Copeland:2006wr}
E.J.~Copeland, M.~Sami and S.~Tsujikawa, \emph{{Dynamics of dark energy}}, \href{https://doi.org/10.1142/S021827180600942X}{\emph{Int. J. Mod. Phys. D} {\bfseries 15} (2006) 1753} [\href{https://arxiv.org/abs/hep-th/0603057}{{\ttfamily hep-th/0603057}}].

\bibitem{Li:2011sd}
M.~Li, X.-D.~Li, S.~Wang and Y.~Wang, \emph{{Dark Energy}}, \href{https://doi.org/10.1088/0253-6102/56/3/24}{\emph{Commun. Theor. Phys.} {\bfseries 56} (2011) 525} [\href{https://arxiv.org/abs/1103.5870}{{\ttfamily 1103.5870}}].

\bibitem{Carroll:1998zi}
S.M.~Carroll, \emph{{Quintessence and the rest of the world}}, \href{https://doi.org/10.1103/PhysRevLett.81.3067}{\emph{Phys. Rev. Lett.} {\bfseries 81} (1998) 3067} [\href{https://arxiv.org/abs/astro-ph/9806099}{{\ttfamily astro-ph/9806099}}].

\bibitem{Bertotti:2003rm}
B.~Bertotti, L.~Iess and P.~Tortora, \emph{{A test of general relativity using radio links with the Cassini spacecraft}}, \href{https://doi.org/10.1038/nature01997}{\emph{Nature} {\bfseries 425} (2003) 374}.

\bibitem{Schoneberg:2021qvd}
N.~Sch\"oneberg, G.~Franco~Abell\'an, A.~P\'erez~S\'anchez, S.J.~Witte, V.~Poulin and J.~Lesgourgues, \emph{{The H0 Olympics: A fair ranking of proposed models}}, \href{https://doi.org/10.1016/j.physrep.2022.07.001}{\emph{Phys. Rept.} {\bfseries 984} (2022) 1} [\href{https://arxiv.org/abs/2107.10291}{{\ttfamily 2107.10291}}].

\bibitem{DiValentino:2021izs}
E.~Di~Valentino, O.~Mena, S.~Pan, L.~Visinelli, W.~Yang, A.~Melchiorri et~al., \emph{{In the realm of the Hubble tension\textemdash{}a review of solutions}}, \href{https://doi.org/10.1088/1361-6382/ac086d}{\emph{Class. Quant. Grav.} {\bfseries 38} (2021) 153001} [\href{https://arxiv.org/abs/2103.01183}{{\ttfamily 2103.01183}}].

\bibitem{DiValentino:2017iww}
E.~Di~Valentino, A.~Melchiorri and O.~Mena, \emph{{Can interacting dark energy solve the $H_0$ tension?}}, \href{https://doi.org/10.1103/PhysRevD.96.043503}{\emph{Phys. Rev. D} {\bfseries 96} (2017) 043503} [\href{https://arxiv.org/abs/1704.08342}{{\ttfamily 1704.08342}}].

\bibitem{DiValentino:2019ffd}
E.~Di~Valentino, A.~Melchiorri, O.~Mena and S.~Vagnozzi, \emph{{Interacting dark energy in the early 2020s: A promising solution to the $H_0$ and cosmic shear tensions}}, \href{https://doi.org/10.1016/j.dark.2020.100666}{\emph{Phys. Dark Univ.} {\bfseries 30} (2020) 100666} [\href{https://arxiv.org/abs/1908.04281}{{\ttfamily 1908.04281}}].

\bibitem{Yang:2021hxg}
W.~Yang, S.~Pan, E.~Di~Valentino, O.~Mena and A.~Melchiorri, \emph{{2021-H0 odyssey: closed, phantom and interacting dark energy cosmologies}}, \href{https://doi.org/10.1088/1475-7516/2021/10/008}{\emph{JCAP} {\bfseries 10} (2021) 008} [\href{https://arxiv.org/abs/2101.03129}{{\ttfamily 2101.03129}}].

\bibitem{Honorez_2010}
L.L.~Honorez, B.A.~Reid, O.~Mena, L.~Verde and R.~Jimenez, \emph{Coupled dark matter-dark energy in light of near universe observations}, \href{https://doi.org/10.1088/1475-7516/2010/09/029}{\emph{Journal of Cosmology and Astroparticle Physics} {\bfseries 2010} (2010) 029–029}.

\bibitem{Gavela:2010tm}
M.B.~Gavela, L.~Lopez~Honorez, O.~Mena and S.~Rigolin, \emph{{Dark Coupling and Gauge Invariance}}, \href{https://doi.org/10.1088/1475-7516/2010/11/044}{\emph{JCAP} {\bfseries 11} (2010) 044} [\href{https://arxiv.org/abs/1005.0295}{{\ttfamily 1005.0295}}].

\bibitem{He:2008si}
J.-H.~He, B.~Wang and E.~Abdalla, \emph{{Stability of the curvature perturbation in dark sectors' mutual interacting models}}, \href{https://doi.org/10.1016/j.physletb.2008.11.062}{\emph{Phys. Lett. B} {\bfseries 671} (2009) 139} [\href{https://arxiv.org/abs/0807.3471}{{\ttfamily 0807.3471}}].

\bibitem{Valiviita:2008iv}
J.~Valiviita, E.~Majerotto and R.~Maartens, \emph{{Instability in interacting dark energy and dark matter fluids}}, \href{https://doi.org/10.1088/1475-7516/2008/07/020}{\emph{JCAP} {\bfseries 07} (2008) 020} [\href{https://arxiv.org/abs/0804.0232}{{\ttfamily 0804.0232}}].

\bibitem{Yang:2017ccc}
W.~Yang, S.~Pan and D.F.~Mota, \emph{{Novel approach toward the large-scale stable interacting dark-energy models and their astronomical bounds}}, \href{https://doi.org/10.1103/PhysRevD.96.123508}{\emph{Phys. Rev. D} {\bfseries 96} (2017) 123508} [\href{https://arxiv.org/abs/1709.00006}{{\ttfamily 1709.00006}}].

\bibitem{Gavela:2009cy}
M.B.~Gavela, D.~Hernandez, L.~Lopez~Honorez, O.~Mena and S.~Rigolin, \emph{{Dark coupling}}, \href{https://doi.org/10.1088/1475-7516/2009/07/034}{\emph{JCAP} {\bfseries 07} (2009) 034} [\href{https://arxiv.org/abs/0901.1611}{{\ttfamily 0901.1611}}].

\bibitem{Yang:2022csz}
W.~Yang, S.~Pan, O.~Mena and E.~Di~Valentino, \emph{{On the dynamics of a dark sector coupling}}, \href{https://doi.org/10.1016/j.jheap.2023.09.001}{\emph{JHEAp} {\bfseries 40} (2023) 19} [\href{https://arxiv.org/abs/2209.14816}{{\ttfamily 2209.14816}}].

\bibitem{Yang:2019uzo}
W.~Yang, O.~Mena, S.~Pan and E.~Di~Valentino, \emph{{Dark sectors with dynamical coupling}}, \href{https://doi.org/10.1103/PhysRevD.100.083509}{\emph{Phys. Rev. D} {\bfseries 100} (2019) 083509} [\href{https://arxiv.org/abs/1906.11697}{{\ttfamily 1906.11697}}].

\bibitem{Yang:2020uga}
W.~Yang, E.~Di~Valentino, O.~Mena, S.~Pan and R.C.~Nunes, \emph{{All-inclusive interacting dark sector cosmologies}}, \href{https://doi.org/10.1103/PhysRevD.101.083509}{\emph{Phys. Rev. D} {\bfseries 101} (2020) 083509} [\href{https://arxiv.org/abs/2001.10852}{{\ttfamily 2001.10852}}].

\bibitem{Nunes:2022bhn}
R.C.~Nunes, S.~Vagnozzi, S.~Kumar, E.~Di~Valentino and O.~Mena, \emph{{New tests of dark sector interactions from the full-shape galaxy power spectrum}}, \href{https://doi.org/10.1103/PhysRevD.105.123506}{\emph{Phys. Rev. D} {\bfseries 105} (2022) 123506} [\href{https://arxiv.org/abs/2203.08093}{{\ttfamily 2203.08093}}].

\bibitem{vandeBruck:2022xbk}
C.~van~de Bruck, G.~Poulot and E.M.~Teixeira, \emph{{Scalar field dark matter and dark energy: a hybrid model for the dark sector}}, \href{https://doi.org/10.1088/1475-7516/2023/07/019}{\emph{JCAP} {\bfseries 07} (2023) 019} [\href{https://arxiv.org/abs/2211.13653}{{\ttfamily 2211.13653}}].

\bibitem{DiValentino:2019jae}
E.~Di~Valentino, A.~Melchiorri, O.~Mena and S.~Vagnozzi, \emph{{Nonminimal dark sector physics and cosmological tensions}}, \href{https://doi.org/10.1103/PhysRevD.101.063502}{\emph{Phys. Rev. D} {\bfseries 101} (2020) 063502} [\href{https://arxiv.org/abs/1910.09853}{{\ttfamily 1910.09853}}].

\bibitem{Giare:2024ytc}
W.~Giar\`e, Y.~Zhai, S.~Pan, E.~Di~Valentino, R.C.~Nunes and C.~van~de Bruck, \emph{{Tightening the reins on non-minimal dark sector physics: Interacting Dark Energy with dynamical and non-dynamical equation of state}},  \href{https://arxiv.org/abs/2404.02110}{{\ttfamily 2404.02110}}.

\bibitem{Carrillo_Gonz_lez_2018}
M.~Carrillo~González and M.~Trodden, \emph{Field theories and fluids for an interacting dark sector}, \href{https://doi.org/10.1103/physrevd.97.043508}{\emph{Physical Review D} {\bfseries 97} (2018) }.

\bibitem{Johnson:2020gzn}
J.P.~Johnson and S.~Shankaranarayanan, \emph{{Cosmological perturbations in the interacting dark sector: Mapping fields and fluids}}, \href{https://doi.org/10.1103/PhysRevD.103.023510}{\emph{Phys. Rev. D} {\bfseries 103} (2021) 023510} [\href{https://arxiv.org/abs/2006.04618}{{\ttfamily 2006.04618}}].

\bibitem{Amendola:1999er}
L.~Amendola, \emph{{Coupled quintessence}}, \href{https://doi.org/10.1103/PhysRevD.62.043511}{\emph{Phys. Rev. D} {\bfseries 62} (2000) 043511} [\href{https://arxiv.org/abs/astro-ph/9908023}{{\ttfamily astro-ph/9908023}}].

\bibitem{Blas:2011rf}
D.~Blas, J.~Lesgourgues and T.~Tram, \emph{{The Cosmic Linear Anisotropy Solving System (CLASS) II: Approximation schemes}}, \href{https://doi.org/10.1088/1475-7516/2011/07/034}{\emph{JCAP} {\bfseries 07} (2011) 034} [\href{https://arxiv.org/abs/1104.2933}{{\ttfamily 1104.2933}}].

\bibitem{Hwang_2009}
J.-c.~Hwang and H.~Noh, \emph{Axion as a cold dark matter candidate}, \href{https://doi.org/10.1016/j.physletb.2009.08.031}{\emph{Physics Letters B} {\bfseries 680} (2009) 1–3}.

\bibitem{Poulin_2018}
V.~Poulin, T.L.~Smith, D.~Grin, T.~Karwal and M.~Kamionkowski, \emph{Cosmological implications of ultralight axionlike fields}, \href{https://doi.org/10.1103/physrevd.98.083525}{\emph{Physical Review D} {\bfseries 98} (2018) }.

\bibitem{Christopherson_2009}
A.J.~Christopherson and K.A.~Malik, \emph{The non-adiabatic pressure in general scalar field systems}, \href{https://doi.org/10.1016/j.physletb.2009.04.003}{\emph{Physics Letters B} {\bfseries 675} (2009) 159–163}.

\bibitem{Koshelev_2011}
N.~Koshelev, \emph{Non-adiabatic perturbations in multi-component perfect fluids}, \href{https://doi.org/10.1088/1475-7516/2011/04/021}{\emph{Journal of Cosmology and Astroparticle Physics} {\bfseries 2011} (2011) 021–021}.

\bibitem{Caldera-Cabral:2009hoy}
G.~Caldera-Cabral, R.~Maartens and B.M.~Schaefer, \emph{{The Growth of Structure in Interacting Dark Energy Models}}, \href{https://doi.org/10.1088/1475-7516/2009/07/027}{\emph{JCAP} {\bfseries 07} (2009) 027} [\href{https://arxiv.org/abs/0905.0492}{{\ttfamily 0905.0492}}].

\bibitem{Zhao:2022ycr}
Y.~Zhao, Y.~Liu, S.~Liao, J.~Zhang, X.~Liu and W.~Du, \emph{{Constraining interacting dark energy models with the halo concentration\textendash{}mass relation}}, \href{https://doi.org/10.1093/mnras/stad1814}{\emph{Mon. Not. Roy. Astron. Soc.} {\bfseries 523} (2023) 5962} [\href{https://arxiv.org/abs/2212.02050}{{\ttfamily 2212.02050}}].

\bibitem{Zhai:2023yny}
Y.~Zhai, W.~Giar\`e, C.~van~de Bruck, E.~Di~Valentino, O.~Mena and R.C.~Nunes, \emph{{A consistent view of interacting dark energy from multiple CMB probes}}, \href{https://doi.org/10.1088/1475-7516/2023/07/032}{\emph{JCAP} {\bfseries 07} (2023) 032} [\href{https://arxiv.org/abs/2303.08201}{{\ttfamily 2303.08201}}].

\bibitem{Ilic:2020onu}
S.~Ili\'c, M.~Kopp, C.~Skordis and D.B.~Thomas, \emph{{Dark matter properties through cosmic history}}, \href{https://doi.org/10.1103/PhysRevD.104.043520}{\emph{Phys. Rev. D} {\bfseries 104} (2021) 043520} [\href{https://arxiv.org/abs/2004.09572}{{\ttfamily 2004.09572}}].

\bibitem{Pan:2022qrr}
S.~Pan, W.~Yang, E.~Di~Valentino, D.F.~Mota and J.~Silk, \emph{{IWDM: the fate of an interacting non-cold dark matter \textemdash{} vacuum scenario}}, \href{https://doi.org/10.1088/1475-7516/2023/07/064}{\emph{JCAP} {\bfseries 07} (2023) 064} [\href{https://arxiv.org/abs/2211.11047}{{\ttfamily 2211.11047}}].

\bibitem{khurshudyan2024constraints}
M.~Khurshudyan and E.~Elizalde, \emph{Constraints on prospective deviations from the cold dark matter model using a gaussian process},  2024.

\bibitem{Brout:2022vxf}
D.~Brout et~al., \emph{{The Pantheon+ Analysis: Cosmological Constraints}}, \href{https://doi.org/10.3847/1538-4357/ac8e04}{\emph{Astrophys. J.} {\bfseries 938} (2022) 110} [\href{https://arxiv.org/abs/2202.04077}{{\ttfamily 2202.04077}}].

\bibitem{DESI:2024mwx}
{\scshape DESI} collaboration, \emph{{DESI 2024 VI: Cosmological Constraints from the Measurements of Baryon Acoustic Oscillations}},  \href{https://arxiv.org/abs/2404.03002}{{\ttfamily 2404.03002}}.

\bibitem{Lewis:1999bs}
A.~Lewis, A.~Challinor and A.~Lasenby, \emph{{Efficient computation of CMB anisotropies in closed FRW models}}, \href{https://doi.org/10.1086/309179}{\emph{Astrophys. J.} {\bfseries 538} (2000) 473} [\href{https://arxiv.org/abs/astro-ph/9911177}{{\ttfamily astro-ph/9911177}}].

\end{thebibliography}\endgroup

\end{document}